\documentclass[useAMS,usenatbib]{mn2e}

\usepackage{graphicx}
\usepackage{amssymb}

\newcommand{\apj}{ApJ}

\newcommand{\aap}{A\&A}

\newcommand{\aj}{AJ}
\newcommand{\mnras}{MNRAS}

\newcommand{\MC}{\multicolumn}
\newcommand{\kms}{km~s$^{-1}$}

\newcommand{\sunn}{$_{\odot}$}
\DeclareRobustCommand{\ion}[2]{%
\relax\ifmmode
\ifx\testbx\f
{\mathrm{#1\,\textsc{#2}}}\else
{\mathrm{#1\,\mathsc{#2}}}\fi
\else\textup{#1\,{\mdseries\textsc{#2}}}%
\fi}

\title[Ionized gas kinematics in nine XMD galaxies]{Very metal-poor galaxies:
ionized gas kinematics in nine
objects\thanks{Based on observations obtained with the Special Astrophysical
Observatory RAS 6-m telescope.}}
\author[A. V. Moiseev,  S. A. Pustilnik, A. Y. Kniazev]
{A. V. Moiseev,$^1$$\thanks{moisav@gmail.com(AVM),sap@sao.ru(SAP),akniazev@saao.ac.za(AYK)}$
S. A. Pustilnik,$^{1,4}$
A. Y. Kniazev$^{2,3}$        \\
$^1$ Special Astrophysical Observatory of RAS, Nizhnij Arkhyz,
  Karachai-Circassia 369167, Russia\\
$^2$ South African Astronomical Observatory, PO Box 9, 7935 Observatory,
  Cape Town, South Africa\\
$^3$ Southern African Large Telescope Foundation, PO Box 9, 7935 Observatory,
   Cape Town, South Africa\\
$^4$ I.Newton Institute, Chile, SAO branch, Nizhnij Arkhyz, Russia}

\begin{document}

\date{Accepted March ?? 2010. Received June NN, 2009}

\pagerange{\pageref{firstpage}--\pageref{lastpage}} \pubyear{2010}

\maketitle

\label{firstpage}

\begin{abstract}

The study of ionized gas morphology and kinematics in nine eXtremely
Metal-Deficient (XMD) galaxies with the scanning Fabry-Perot interferometer
on the SAO 6-m telescope is presented. Some of these very rare objects
(with currently known range of O/H of 7.12 $<$ 12+$\log$(O/H) $<$7.65, or
Z\sunn/35 $<$ Z $<$Z\sunn/10) are believed to be the best proxies of `young'
low-mass galaxies in the high-redshift Universe.
One of the main goals of this study is to look for possible evidence
of star formation (SF) activity induced by external perturbations.
Recent results from HI mapping of a small subsample of XMD star-forming
galaxies provided confident evidence for the important role of
interaction-induced SF. Our observations provide complementary or new
information that the great majority of the studied XMD dwarfs  have
strongly disturbed gas morphology and kinematics or
the presence of detached components. We approximate the observed velocity
fields by simple models of a rotating tilted thin disc, which allow
us the robust detection of non-circular gas motions.
These data, in turn, indicate the important role of current/recent
interactions and mergers in the observed enhanced star formation.
As a by-product of our observations, we obtained data for two LSB dwarf
galaxies: Anon~J012544+075957 that is a companion of the merger system
UGC~993, and SAO 0822+3545 which shows off-centre, asymmetric, low
SFR star-forming regions, likely induced by the interaction with the companion
XMD dwarf HS 0822+3542.
\end{abstract}


\section[]{INTRODUCTION}
\label{sec:intro}

The group of very rare dwarf galaxies with gas metallicities Z below
Z\sunn/10  (or 12+$\log$(O/H)$<$7.65, e.g., review by \citet{KO}),
are called eXtremely Metal-Deficient, or XMD galaxies.
Such objects comprise only about two per cent of known blue compact
galaxies (BCGs), low-mass galaxies with active star formation
\citep[e.g.][]{HSS6}.
The latter, in turn, comprise about 6 per cent of the entire  local
dwarf galaxy
population  \citep[][ and references therein]{Lee09}.
Thanks to intensive searches for new XMD galaxies in several recent large
surveys,  \citep*[e.g.,][ among others, and references therein]
{KISS,HSS-LM,Kiel2002,SDSS_XMP,HSS6,Brown08,Papa08,Guseva09},  the number of
currently known such objects in the local Universe is about 100.
There is also progress in identification of such objects at large
redshifts \citep{Kakazu07}.

The interest in these {\it atypical}, local Universe galaxies is
caused by
the similarity of their properties to low-mass galaxies in the early
Universe,
when the typical metallicity of galaxies was one-two orders of magnitude
smaller than the solar one.
Therefore, understanding the specific processes of star
formation (SF) and evolution in nearby galaxies with such low metallicity
will provide insights into similar processes in young galaxies at high
redshifts. To better understand various aspects of XMD galaxies star formation and
evolution, we conduct an extensive study of a subsample of XMD galaxies
in optical, NIR and radio domains
\citep*[e.g.,][]{SAO0822,HS0837,SBS0335,DDO68,PM07,Ekta08,DDO68_sdss,
SBS0335_GMRT}.

One of the fundamental questions of SF in low-mass, gas-rich galaxies
is the trigger mechanisms of starbursts: whether the intrinsic
instabilities in gas discs are capable to produce the observed level of
enhanced (bursting) SF (e.g., models by \citet{Pelupessy04} and
\citet{dimatteo2008};  observations of Virgo cluster dIrs by
\citet{Brosch04} and references therein), or external mechanisms
(such as tidal disturbances and mergers) are necessary in a significant
fraction of observed starbursts.

From the analysis of the density distribution in BCGs in comparison to more
typical late-type dwarfs, \citet{Salzer99} concluded that the former have
a higher gas concentration and, therefore, BCGs represent a special group
of galaxies, more susceptible to internal instabilities and
related starbursts. But,  it is unknown whether this is an inherent
property of BCG progenitors, or this higher gas concentration appeared
due to a minor
merger, or as a result of a recent interaction with a barely detectable
companion. If due to mergers or interactions,
BCG progenitors could be the quite common late-type dwarfs. To disentangle
the possible scenarios  one needs a large, representative sample of
starbursting galaxies with well studied and understood kinematics of both
HI and ionized gas.  This is required to determine the prevalence of
external or internal disturbances.

First statistical indications for the possible importance of interactions to
trigger starbursts in BCG progenitors was presented by \citet{Taylor95} in
their search for HI-rich companions in  a small BCG sample. For a larger BCG
sample, the conclusion on the high fraction of interaction-induced starbursts
was made by \citet[][see also \citet{Noeske01}]{Pustilnik01},
based on
the comparison of observed relative distances and `threshold tidal distances',
at which a perturber galaxy should induce shocks in the gas disc and produce
dissipation of gas angular momentum \citep[as suggested by][]{Icke85}.
While that study was {\it indicative} of the important role of collisions in
{\it starbursts} of gas-rich low-mass galaxies, more elaborate tests
should be used to clear up the real role of external factors in starbursts.
In turn, numerous studies of galaxy morphology and kinematics at
redshifts of up to $\sim$1 \citep[e.g.,][]{Puech06,Yang08} show clear
indications on the importance of external mechanisms on the enhanced SF
in a significantly larger galaxy fraction than at the current epoch.

In many cases, the traces of interactions can be hidden if only images of
stellar components are available. Therefore, gas morphology and kinematics provide a better tracer of
interaction-induced starbursts.  To this end,  HI
mapping is a good way for galaxies with sufficiently large HI flux. For less
gas-rich and/or more distant galaxies, the study of the ionized gas kinematics
can be a good, complementary technique. Besides, this
method provides usually a finer angular resolution and allows comparison of
the global gas kinematics derived from HI mapping with the gas motions
within the region of higher density, around the sites of SF. These kind of
studies, initiated by \citet{Ostlin99,Ostlin01} for a small sample of
luminous BCGs, already provide good evidence for merger-induced SF in
such objects. Some indications for the same phenomenon have been
observed by \citet{Garcia08} for a group of five other luminous BCGs and
by \citet{Perez09} for representatives of the similar group of Luminous
Compact Blue Galaxies.

The starbursting XMD galaxies are especially interesting in this aspect,
since the results may have implications for
cosmological young galaxies in the high-redshift Universe.
The evidence of significantly disturbed optical morphology and
interactions,  indicating an unrelaxed state of many XMD galaxies, was already
noticed  by us \citep[e.g.,][]{SBS0335,DDO68,HS2134} and
more recently by \citet{Papa08}. How does this relate to the origin of
XMD galaxies?

There are several options for evolutionary scenarios of XMD dwarf
galaxies, already emphasized in the literature \citep[e.g.,][]{DDO68}.
Two of them relate to gas-rich discs, which are very stable locally,
and being isolated, have been evolving very slowly (like LSB galaxies).
Some of them might even remain dark objects \citep[e.g.,][]{dark07}
(and, thus, `pristine'
like protogalaxies at epochs of the major galaxy formation). Such galaxies
would be very difficult to discover/recognise before they (by chance) have
experienced a strong external perturbation due to an interaction
with  a galaxy-sized mass. When this happens, such galaxies can lose their
global stability and experience a substantial SF burst.
Due to their slow evolution,
during the starburst they would appear as
gas-rich and very metal-poor (XMD) objects. Therefore, the study of gas
kinematics for the sample of XMD galaxies can shed light on their origin as
a group, or can give clues to the diversity of their properties and their
finer classification.

In this paper we present the results of an H$\alpha$ study of the ionized
gas kinematics in nine XMD starbursting galaxies (BCGs)
conducted with the SAO 6-m telescope's scanning Fabry-Perot Interferometer
(FPI).
These galaxies constitute a part of a larger XMD galaxy sample, for
which we conduct a multiwavelength  study of their properties. Most
were selected for this FPI H$\alpha$ study due to their unusual or
disturbed optical morphology. For a fraction of them, the HI mapping data are
available,
which provide an opportunity of the combined view on the gas kinematics.
We plan to continue this project for a larger sample of XMD starbursting
galaxies in order to improve statistics and to search for possible
differences in the properties of the ionized gas kinematics.

Due to volume limitations, the scope of this paper is kept mainly to
observational results for the studied XMD galaxy subsample and their
preliminary analysis. The more advanced analysis, involving multiwavelength
data and new methods of modelling
\citep[like suggested by][]{Barnes2009},  will be presented for individual
galaxies in forthcoming papers. This  paper is organised as follows:
in Sect.~\ref{sec:obs} we describe observations and
their reduction. Sect.~\ref{sec:results} presents the description of data
analysis  and the summary of results. In Sect.~\ref{sec:dis} the
kinematic properties are discussed in more detail and are compared
with other data on these BCGs, and their likely interpretation is suggested.
Sect.~\ref{sec:summ} summarises conclusions of this study.

\section[]{OBSERVATIONS AND DATA REDUCTION}
\label{sec:obs}

\begin{table}
\begin{center}
\caption{Log of the 6-m telescope FPI observations}
\label{t:journal}
\begin{tabular}{llrc} \\ \hline \hline
\MC{1}{c}{ Object }     &
\MC{1}{c}{ Date }       &
\MC{1}{c}{ Exposure }   &
\MC{1}{c}{  ang. resol. }     \\

\MC{1}{c}{ }       &
\MC{1}{c}{ }       &
\MC{1}{c}{ time (s) }    &
\MC{1}{c}{ (arcsec)}    \\

\MC{1}{c}{ (1) } &
\MC{1}{c}{ (2) } &
\MC{1}{c}{ (3) } &
\MC{1}{c}{ (4) }  \\
\hline
\\[-0.3cm]
SDSS~J0113+0052  & 11.01.07    & 36x200   &  4.0  \\
HS~0122+0743     & 10.01.07    & 2x36x90  &  2.3  \\
SBS~0335-052W    & 14.01.08    & 2x36x150 &  1.4  \\
SBS~0335-052E    & 14.01.08    & 2x36x150 &  1.4  \\
HS~0822+3542     & 10.01.07    & 36x250   &  1.4  \\
SDSS~J1044+0353  & 14.01.08    & 36x150   &  1.4  \\
SBS~1116+517     & 10.01.07    & 36x250   &  1.5  \\
SBS~1159+545     & 14.01.08    & 36x170   &  1.8  \\
HS~2236+1344     & 08.09.05    & 36x250   &  1.0  \\
\hline \hline \\[-0.2cm]
\end{tabular}
\end{center}
\end{table}

The observations of XMD galaxies
were conducted with the multimode instrument SCORPIO \citep{SCORPIO}
installed in the prime focus of the SAO 6-m telescope (BTA)
during 3 runs (see Table~\ref{t:journal} for details and Table~\ref{t:list}
for galaxy main parameters).
The desired spectral interval in the neighbourhood of the redshifted
H$\alpha$ line was
cut using a set of narrow-band filters (\mbox{$FWHM=15-20$\AA}).
The width of the free spectral interval between the neighbouring orders of
interference was equal to 13\,\AA\, (about 600~\kms). The resolution of the
interferometer ($FWHM$ of the instrumental profile) was 0.8~\AA\,
(35~\kms) for a 0.36\AA/channel sampling. The detector was a 2048$\times$2048
EEV 42-40 CCD operating in the 2$\times$2   and 4$\times$4 (for
SDSS~J0113+0052=UGC~772  and HS~0122+0743) pixel instrumental averaging mode
in order to reduce the readout time. The corresponding image sampling was
0\farcs36  and 0\farcs72 pixel$^{-1}$ respectively. The resulting field of
view has a size of 6\farcm1. In column 4 we also present the final
angular resolution of our data which is a combination of
the real seeing and the data reduction smoothing.

We took a total of 36 successive interferograms per object with
different gaps between the plates of the FPI.  The observational
data were reduced using an IDL-based software package \citep{Moiseev02,ME08}.
After primary reduction, night-sky line subtraction, and the wavelength
calibration, the observations were reduced to the form of a data cube,
where each pixel in the  field of view contains a 36-channel spectrum.
We removed the ghost images that appear in the plates of the  interferometer
using the algorithms
described by \citet{ME08}. To check the quality of ghosts removal on the data
of extended objects, we  observed HS~0122+0743 and SBS~0335--052E,W
successively in two fields  turned in the position angle.

We fitted the  H$\alpha$ emission profiles to the Voigt function,
which in most cases describes the observed contour fairly well
\citep[see discussion in][]{ME08}. The profile fitting results were used
to construct the two-dimensional field of line-of-sight velocities of
ionized gas, the map of velocities dispersion, and images of the galaxy
in the H$\alpha$ line and in the neighbouring continuum. We estimate the
accuracy of line-of-sight velocity measurements through the measured
S/N ratios, using the relations given in Fig.~5 of \citet{ME08}.
The accuracy was found to be \mbox{1--5~\kms} in the bright HII regions and
amounts to \mbox{10--15~\kms} in the regions with the minimal surface
brightness, where the emission line shows up at the S/N ratio
level of $S/N\approx5$.
In Fig.~\ref{fig:line_prof} we show the examples of representative
H$\alpha$-line profiles for three program galaxies  SBS~0335--052W and E and
SBS~1116+517.

\section[]{RESULTS AND ANALYSIS}
\label{sec:results}

In Table~\ref{t:list} we present the summary of the main observational
parameters of the studied XMD galaxies.  Since the accuracy of
our total H$\alpha$ flux estimates (10--30 per cent) is in general worse than
that for dedicated H$\alpha$ photometry, we give our values only for unknown
H$\alpha$ fluxes. For the remaining objects, we cite data from the literature.
Absolute magnitudes, M$_{\rm B}$, are calculated for distances taken from
NED\footnote{http://nedwww.ipac.caltech.edu/}, except for HS~0822+3542 and its companion LSB dwarf. As discussed in
\citet{J0926}, they are situated in the region with large peculiar
negative radial velocities ($\sim$300~\kms) due to the effect of the huge
Local Void \citep{Tully08}. So to derive their distances, this peculiar
velocity was accounted for.

The dominant or important mode of gas motions in late-type, low-mass isolated
galaxies (even for objects with luminosities corresponding to M$_{\rm V}$
fainter than --14.0) is rotation in a gravitational field of a Dark Matter
(DM) halo and  baryonic disc \citep[e.g.,][ and references therein]
{Begum06,Begum08}. Two types of processes  can significantly affect
the regular velocity field attributed to rotation:
the star formation activity and tidal interactions.

\subsection{Kinematics of SF induced shells}

The strong SF activity  releases a significant amount of kinetic and thermal
energy in short periods, and it results in
the formation of hot bubbles and shells with typical sizes of a
hundred pc to about one kpc. Such shells, with maximal
observed line-of-sight velocities in the range of ten to a hundred \kms\ are
recognised in both long-slit spectra and 2D H$\alpha$ velocity fields of
star-forming galaxies \citep[e.g.,][among others]
{Martin96,Martin98,Zasov00,VV2,Lozinskaya06,Martinez07,Bordalo09}.
In principle, the large-scale velocity fields in such expanding ionized
shells are sufficiently simple.
However, in case of dwarf galaxies, which have relatively small
rotation velocities, shells can significantly perturb the disc velocity
field.
In Appendix~\ref{App1} we present the simulated velocity field of
dwarf galaxies with shells for several simple cases.
Using the characteristic parameters  typical of our sample (disc size,
inclination, rotation curve, etc.), we describe the technique for such shells
recognition and masking  in the observed velocity fields.
Fortunately, in many  simple theoretical scenarios it is possible
to extract the real parameters of global rotation, even when the shell expansion velocity is larger than the
line-of-sight projection of global rotation on given radii.

\subsection{External perturbations of velocity fields}

It is not yet well understood how the strength of starbursts and that of
related expanding bubbles are connected with internal or external triggers.
But, since the external perturbations (e.g., due to galaxy collisions or
extragalactic cloud infall) provide disturbances which are additional
to internal ones, they are naturally expected to trigger the stronger
disturbances in galactic gas and to result in more intense star formation.
Besides, in general, the characteristic timescales of singular star-forming
events in isolated low-mass galaxies (of ten to several tens Myr)
are shorter than those for the events induced by external interactions.
The latter are comparable to rotation periods, and usually are of the order
of a hundred Myr and more, the typical timescale for galaxy response to a
collision-induced perturbation \citep[e.g.,][and references therein]
{dimatteo2008}.

This is due to different spatial scales involved in such SF episodes.
For internal perturbations we usually have some local enhancement, while
for external ones the gas flows on scales of the whole galactic disc are
involved. Therefore, one can expect rather different
effects in isolated and interacting starbursting galaxies. In the former,
the gas velocity field should display the characteristic rotation pattern
with rather localised disturbances due to the starburst-related shells.
More or less typical dIrr galaxies can serve as templates for such kinematics.
In the interaction-induced starbursts one can
observe, besides the overall rotation and the appearance of shell kinematics, the effects of
tidally disturbed velocity fields to various degrees.
In extreme cases of a major merger, the strong disturbance of the
initial equilibrium rotation field can almost wash out the regular component.

Apart from non-destructing (fly-by) collisions, various types of mergers
(major,
intermediate mass-ratio, and minor [M$_{1}$:M$_{2}$ $>$ 5]) can trigger
late-type galaxy starbursts. Depending on the stage and mass proportion,
the velocity fields can show various patterns -- from very  disturbed
with no traces of regular rotation to more or less regular rotation
with residual motions (e.g. counter-rotation in the central region,
where  a smaller component has sunk as the result of minor merger).
As recent simulations demonstrate \citep[e.g.,][]{Kronberger06,Pedrosa08},
the bifurcations in velocity curves should appear (that
is prominent asymmetries in receding and approaching branches of rotation
curves) as a result of close recent encounters. Various appearances of
complex velocity fields resulting from major mergers of disc galaxies were
demonstrated by \citet{Jesseit07}.

\subsection{Fitting results}

From the above discourse it is clear that
it is quite natural to analyse  H$\alpha$ velocity fields in
the studied XMD dwarf galaxies by trying to fit them with a model of a flat
rotating disc and to compare the residual velocity field against various
patterns characteristic of shells, warps, central counter-rotation,
discontinuities, etc. Of course, the morphology of both stellar continuum
and H$\alpha$ emission should be taken into account in the course of the
results' interpretation.

We analysed the velocity fields using the well-known  approximation of a
thin rotating disc ('tilted-rings' method), which is generally accepted
in analysis of gas kinematics.
For the method  description and references to the original works, see
 \citet{Moiseev2004}. This method calculates the disc
orientation parameters (major axis position angle $PA$, inclination $i$,
systemic velocity $V_{sys}$) together with radial variations of the rotation
velocity $V_{rot}$. In special cases of non-circular motions,  we can also
estimate radial dependencies of $PA$ and $V_{sys}$.

In Table \ref{t:fit_disc} the `best-fitting' model parameters are summarised:
systemic velocity V$_{\rm sys}$, position angle $PA$ and inclination angle
$i$, along with relevant brief comments. Unfortunately, from the kinematic
model we could only derive $i$ for three galaxies, whose inclinations are
indicated
with errors in the Table~\ref{t:fit_disc}. For J0113+0052 and SAO~0822+3545,
we adopted
the inclinations from published HI maps. In the
remaining cases, to estimate the inclination angle we adopted the approach
based on the morphology of the overall H$\alpha$ emission and the
observed ratio of minor to major axis of H$\alpha$ `ellipsoid' $p$=$a/b$.
Assuming that this reflects the real gas distribution for an inclined
`thick' (typical of dwarf galaxies) disc with the intrinsic axial
ratio $q$, we estimated the
inclination $i$ with well-known formula $[cos(i)]^2 = (p^2-q^2)/(1-q^2)$.
The estimate of parameter $q$ was adopted according to formula A6 from
\citet{staveley92}.
We understand that in many cases, where we infer strong disturbances in
morphology and velocity field, the above assumption can be hardly justified,
and hence the adopted inclination angle and the related amplitude of rotation
velocity can be rather uncertain. But the concrete choice of $i$ does not
affect the overall radial dependence of the model rotation velocity.
A more detailed description of the data and their model fits, along with
discussion of relevant kinematics and morphology in HI-line and other
points are presented for each individual galaxy in Section~\ref{sec:dis}.

\footnotesize{
\begin{table*}
\caption[]{Parameters of nine XMD and two by-product (below the line) galaxies}
\label{t:list}
\begin{tabular}{lcrrccclrrl} \hline \hline
\\[-0.3cm]
\multicolumn{1}{c}{IAU name }&
\multicolumn{2}{c}{Coord. (2000.0)} &
\multicolumn{1}{c}{V$_{\rm hel}$$\P$} &  \multicolumn{1}{c}{$B_{\rm tot}$$\ddagger$} &
\multicolumn{1}{c}{$M_{\rm B}^{0}$$^*$} &  \multicolumn{1}{l}{O/H$^{\dagger}$} &
\multicolumn{1}{l}{F(H$\alpha$)$^{\S}$} & \multicolumn{1}{l}{$D^{\S\S}$} &
\multicolumn{1}{l}{Scale} &\multicolumn{1}{l}{Alternative}  \\ \cline{2-3}
  & R.A. & Dec. & km $s^{-1}$   & mag  & mag &   &  & Mpc & pc\,arcsec$^{-1}$ & name  \\
  &\ \ $^h$\ \ $^m$\ \ $^s$ \ &  $^{\circ}$\ \ $'$\ \ $''$ &   &   &  &  &   &  &  &  \\
\MC{1}{c}{(1)} & \MC{1}{c}{(2)} & \MC{1}{c}{(3)} & \MC{1}{c}{(4)} & \MC{1}{c}{(5)} & \MC{1}{c}{(6)}      & \MC{1}{c}{(7)} & \MC{1}{c}{(8)} & \MC{1}{c}{(9)} & \MC{1}{c}{(10)} & \MC{1}{c}{(11)}\\ \hline
SDSS J0113+0052      & 01 13 39.40&  +00 52 27.9 &  1176$\pm$~1   & 16.28$^{1}$ &--14.70&7.24 &~6.8      &16.3 & 79 &UGC~772   \\
HS 0122+0743         & 01 25 34.18&  +07 59 22.2 &  2940$\pm$~2   & 15.48$^{2}$ &--17.76&7.63 &20.0$^{2}$&40.3 &195 &UGC 993   \\
SBS 0335-052W        & 03 37 38.46& --05 02 36.3 &  4038$\pm$~2   & 19.14$^{4}$ &--14.73&7.12 &~1.4$^{4}$&53.8 &261 &          \\
SBS 0335-052E        & 03 37 44.04& --05 02 37.6 &  4053$\pm$~2   & 16.95$^{4}$ &--16.92&7.29 &32.0$^{4}$&53.8 &261 &          \\
HS 0822+3542         & 08 25 55.43&  +35 32 31.9 &   727$\pm$~2   & 17.92$^{3}$ &--12.49&7.45 &~7.6$^{6}$&13.5 & 65 &          \\
SDSS J1044+0353      & 10 44 57.84&  +03 53 13.2 &  3865$\pm$~1   & 17.62$^{5}$ &--16.20&7.44 &14.5      &53.8 &261 &          \\
SBS 1116+517         & 11 19 34.29&  +51 30 10.7 &  1342$\pm$~2   & 17.14$^{5}$ &--14.70&7.51 &15.2      &23.1 &112 &Arp's     \\
SBS 1159+545         & 12 02 02.36&  +54 15 50.1 &  3592$\pm$~1   & 19.00$^{5}$ &--14.60&7.49 &~5.5      &52.2 &255 &          \\
HS 2236+1344         & 22 38 31.15&  +14 00 28.6 &  6162$\pm$~2   & 17.88$^{2}$ &--16.90&7.50 &12.0$^{2}$&86.4 &419 &          \\ \hline 
\\[-0.3cm]
Anon J012544+075957  & 01 25 44.18&+07 59 57.0 &  3012$\pm$~3     & 17.0:$^{2}$ &--16.30&     &~1.5      &40.3 &195 &          \\
SAO 0822+3545        & 08 26 05.59&  +35 35 25.7 &   742$\pm$~2   & 17.56$^{3}$ &--12.85&     &~0.4      &13.5 & 65 &          \\
\hline\hline
\\[-0.3cm]
\multicolumn{11}{l}{($\P$) - systemic velocity on our H$\alpha$ data. For UGC~993 this is a mean of V$_{\rm sys}$ for E and W components;} \\
\multicolumn{11}{l}{~~~~~~~for SAO~0822+3545 - this is HI velocity from \citet{Chengalur06}} \\
\multicolumn{11}{l}{($\ddagger$) -- Photometry data are from: $^{1}$ \cite{smoker1996}; $^{2}$ Pustilnik et al. in prep.; $^{3}$ Pustilnik et al. (2003a); } \\
\multicolumn{11}{l}{~~~~~~~$^{4}$ Pustilnik et al. (2004b); $^{5}$ NED/SDSS transformed from $g,r$ to $B$, $^{6}$ \citet{MK06} for H$\alpha$ flux.} \\
\multicolumn{11}{l}{($\dagger$) -- in units 12+$\log$(O/H). Data are from \citet{HSS-LM}, \citet{SDSS_XMP}, \citet{IT07}, } \\
\multicolumn{11}{l}{~~~~~~~\citet{ITG05}.}  \\
\multicolumn{11}{l}{($^*$) -- corrected for A$_{B}$ (according to \citet{Schlegel98}), with distances from Col. 9 } \\
\multicolumn{11}{l}{($^{\S}$) -- in units of 10$^{-14}$ erg~cm$^{-2}$~s$^{-1}$; H$\alpha$ fluxes adopted from literature are coded as photometry data. } \\
\multicolumn{11}{l}{($^{\S\S}$) - from NED [Virgo-inflall; H$_{0}$=73], except HS~0822+3542/SAO~0822+3545, for which additional correction is adopted (see text).}
\end{tabular}
\end{table*}
   }
\normalsize

\footnotesize{
\begin{table*}
\caption[]{Parameters of  model rotating discs for observed galaxies}
\label{t:fit_disc}
\begin{tabular}{lrrll} \hline \hline
\\[-0.3cm]
IAU name      &
V$_{\rm sys}$ &
$PA$(\degr)   &
$i$(\degr)     &
Brief comments \\
\MC{1}{l}{(1)} & \MC{1}{r}{(2)} & \MC{1}{r}{(3)} & \MC{1}{l}{(4)} & \MC{1}{l}{(5)} \\ \hline
J0113+0052  &   1173$\pm$~1   & 199$\pm$5  & $40^*$   & UGC~772. S component is detached.  Probable minor merger      \\
0122+0743W  &   2953$\pm$~2   & 295$\pm$5  & 37$\pm$4 & UGC 993W, component of major merger  in contact     \\
0122+0743E  &   2927$\pm$~2   & 285$\pm$5  & 69       & UGC 993E, component of major merger  in contact      \\
0335--052W  &   4038$\pm$~2   & 64$\pm$8   & 37$\pm$9 & centre between two knots, W comp. of major merger on HI data \\
0335--052E  &   4053$\pm$~2   & $53^*$     & 37       & counter-rotation core, expanding shell, E comp. of major merger on HI data        \\
0822+3542   &    727$\pm$~2   & 51$\pm$7   & 31$\pm$15& HI and H$\alpha$ motions are decoupled    \\
J1044+0353  &   3865$\pm$~1   & 273$\pm$6  & 51       & distorted kinematics of E-half of the disc \\
1116+517    &   1342$\pm$~2   & -22$\pm$7  & 50       & lack of regular rotation,  probable major merging remnant \\
1159+545    &   3592$\pm$~1   & -13$\pm$5  & 38       &  probable merger   \\
2236+1344N  &   6176$\pm$~2   & 165$\pm$30 & 31       & component of major merger in contact  \\
2236+1344S  &   6164$\pm$~1   & 150$\pm$10 & 35       & component of major merger in contact   \\
\hline\\[-0.3cm]
J012544+075957&  3012$\pm$~3  & 324$\pm$~7 & 64       &  probable companion of UGC~993    \\
0822+3545     &   742$\pm$~4  & $105^*$    & $63^*$   & companion of HS~0822+3542.     \\
\hline\hline
\\[-0.3cm]
\multicolumn{5}{l}{($^*$) -- values taken from published HI maps.} \\
\end{tabular}
\end{table*}
      }
\normalsize

\begin{figure}
\includegraphics[width=8.5 cm]{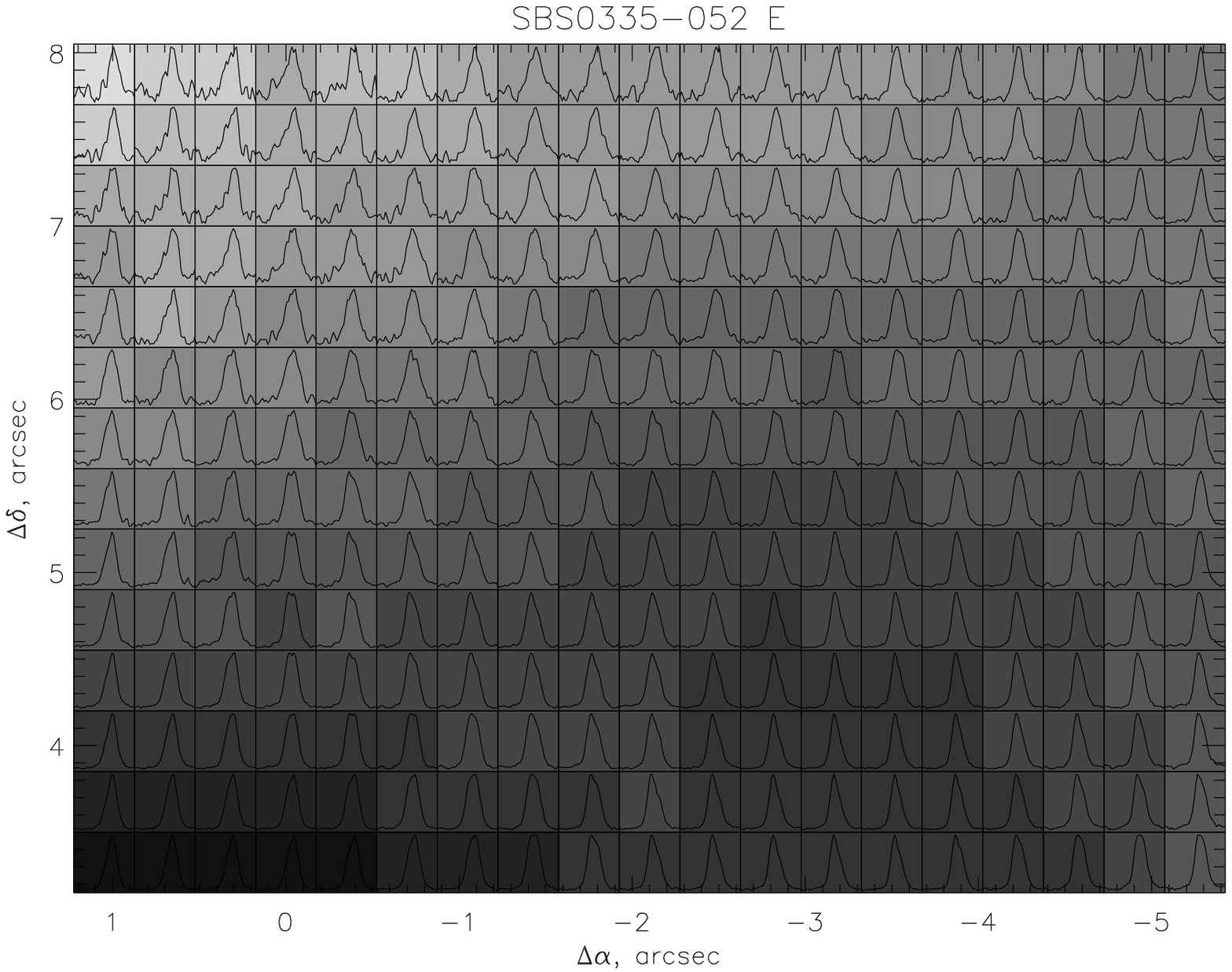}\\
\includegraphics[width=8.5 cm]{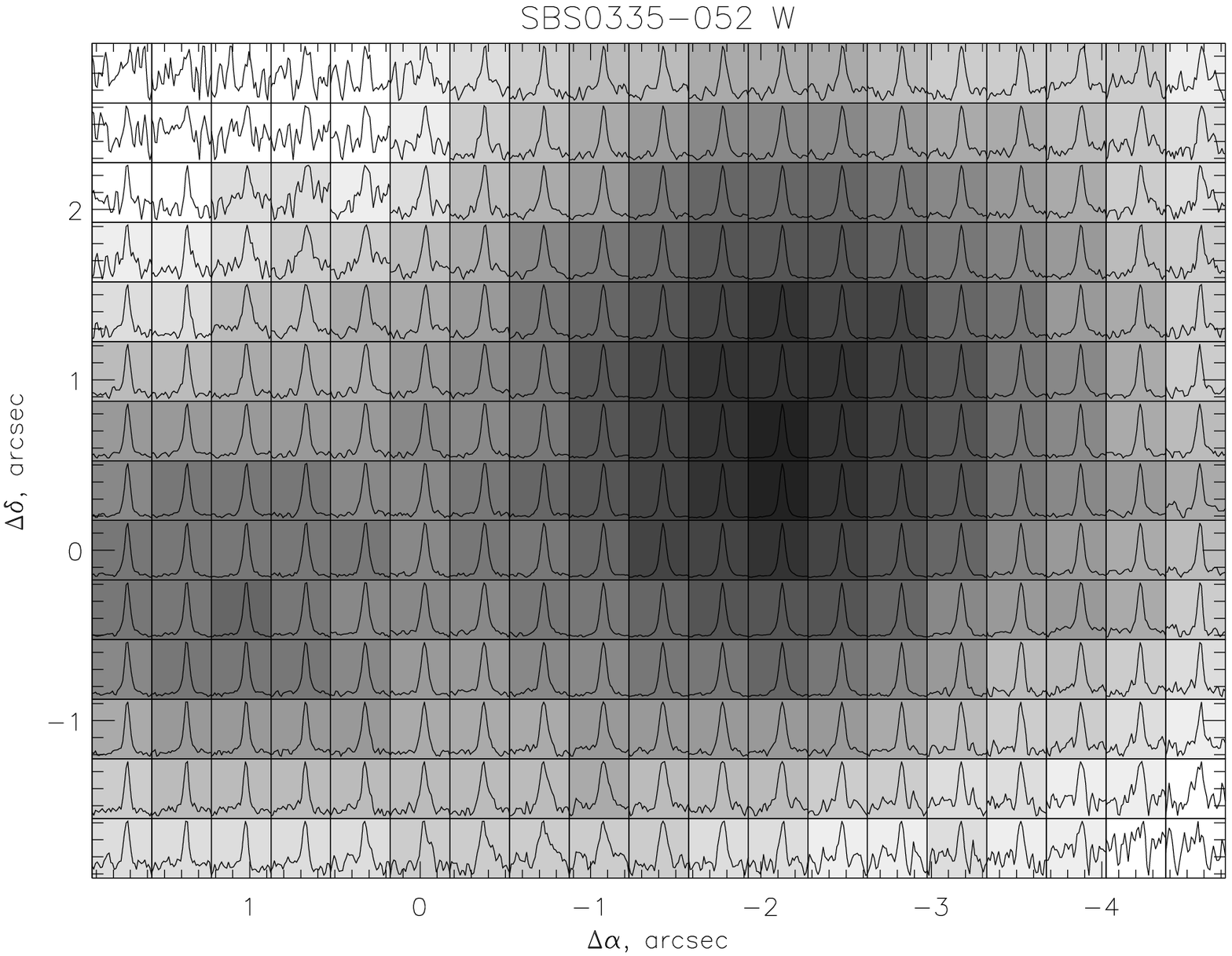}\\
\includegraphics[width=8.5 cm]{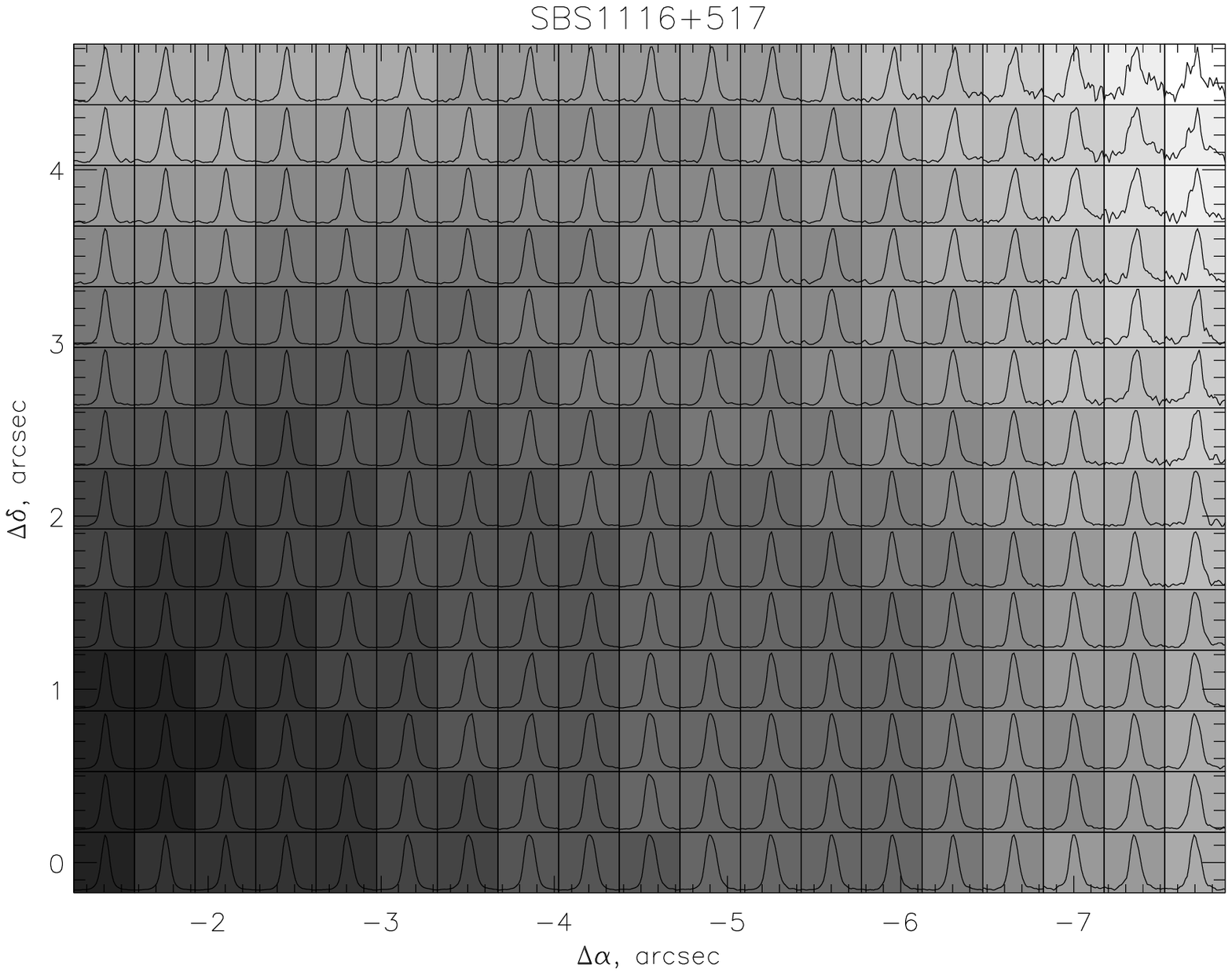}
\caption{Typical individual H$\alpha$-line profiles for each spatial
resolution element, superimposed on the grey-scale brightness distribution
in H$\alpha$-line. The figure demonstrates the range of S/N ratios and
of the line widths. {\bf Top panel: } the region NW of the brightest region in galaxy
SBS~0335--052E. Broadened and double-component structures of H$\alpha$-line
profiles are apparent in the regions where \citet{GIRAFFE} found the evidence
for outflows from the similar line structures. However, our velocity resolution is somewhat insufficient to derive the
reliable parameters of the second component. {\bf Middle panel:}almost the full region of galaxy
SBS~0335--052W. The hints of the line broadening are seen in the outer
LSB regions. {\bf Bottom panel: } the NW part of galaxy SBS~1116+517. There are also
indications on the line broadening at the outermost parts of the nebulosity.}
	\label{fig:line_prof}
\end{figure}

For each of the nine program  XMD galaxies and for two `by-product' LSB
dwarf galaxies, we show the related graphic materials in Figures
\ref{fig:results_p1}-\ref{fig:results_p3}. The colour
images of velocity field data are available in electronic version of the
journal.
They are arranged in columns of 6 images in the following order. The
first (top) panel displays continuum image in $g$-filter from the SDSS or in
the blue band from the digitised Palomar Observatory Sky Survey (POSS). For
SBS~0335--052E,W, the BTA $B$-band images are used from \citet{SBS0335}.
The brightness distribution is shown in logarithmic scale. The second panel
shows the image of integrated emission in the H$\alpha$
line in logarithmic scale. The third panel shows the velocity field, both
in colour and by isovelocity lines.
In the forth panel, we show the `best-fitting' model of thin tilted rotating disc.
The fifth panel presents, in colour, the residual velocity field (observed
minus model) and in the sixth panel the velocity dispersion is shown in
colour. For both panels the intensity of H$\alpha$-line is superimposed by
contours.
The radial dependence of V$_{\rm rot}$ for the `best-fitting'
models of the program galaxies is presented in Fig.~\ref{fig:models}.

\section[]{DISCUSSION}
\label{sec:dis}

The issue of strong interactions of star-forming galaxies in the aspect of
their evolution is actively studied in recent  times \citep[e.g.,][and
references therein]{Puech06,Yang08,dimatteo2008}. Also, many N-body
simulations are performed to allow both
qualitative and quantitative (with more detailed data) classification of
observed velocity fields to originate due to strong interactions or mergers
\citep[e.g.,][]{Kronberger06,Jesseit07}. There is also substantial progress
in morphology analysis to assign galaxies to merger products
\citep[][and references therein]{Conselice03,Lotz08}.

The main goal of these FPI H$\alpha$ study was to
search for the possible appearances of unusual, highly disturbed,
asymmetric velocity fields in the ionized gas kinematics.
The latter might indicate sufficiently strong interactions
or various stages of minor or major mergers. This would allow us to make the
first step to quantify the role of interactions in active SF of XMD galaxies.
Of course, in some cases, the use of only H$\alpha$ kinematics data can
be insufficient to distinguish recent interactions from other types of
kinematic disturbance. However, the combination of ionized gas kinematics
with various morphology indicators, and the use of HI maps (when they are
available), or even of integrated HI profiles \citep{Conselice06}, can lead
to the correct classification.

A complementary study of a subsample of XMD galaxies with similar
goals was conducted based on GMRT HI mapping, and the first results are
presented in \citet{Ekta06,Ekta08}, \citet{Chengalur06}, \citet{SBS0335_GMRT}.
Some earlier results of HI mapping of XMD starbursting galaxies with VLA were
presented by \citet{vZee98} for I~Zw~18 and \citet{PustilnikVLA} for
SBS~0335--052E,W. The H$\alpha$ kinematics of ionized gas in the prototype
XMD galaxy I~Zw~18 was analysed by \citet{Petrosian95}.  All these data
evidence for importance of interactions in the  current starbursts of studied
XMD galaxies.

If we consider the star-forming or blue compact galaxies (BCGs) in general,
there are very few  2D studies of H$\alpha$ kinematics. However, for
the majority of them, the morphology data indicate strong disturbances
in the outer parts. For some of them, such as II~Zw~40, there is clear
evidence of a recent  merger. Several {\it luminous} BCGs were studied
by \citet{Ostlin99}
via imaging in several bands and through H$\alpha$ kinematics with FPI. These
authors concluded that all of their galaxies are best treated as being in the
stage of merging.

It is worth noting, that in such type of studies, the knowledge of the
gas kinematics in the outer parts of a disc can be crucial.
In particular, a good illustration is the luminous BCG Arp~212
(III~Zw~102). For this galaxy, \citet{Garcia08} obtained the ionized gas
velocity field for the circumnuclear region
from integral-field spectroscopy with {\it INTEGRAL} fiber-based  system.
From these spatial limited data, they suggested that the galaxy
``shows a velocity
field resembling a rotational system''. However, the large-scale
H$\alpha$ velocity field taken with FPI, revealed a more complex picture where
the outer HII regions belong to a warped polar ring, formed via the
external gas accretion \citep{Moiseev2008}.
We emphasize that the spatial coverage in our study for all program
galaxies, due to the large SCORPIO field-of-view,
was sufficient to include their distant
periphery. Thus we do not miss any substantial gas velocity field deviations,
as long as they have H$\alpha$ emission above the detection threshold.

 First, we discuss the analysis and model fitting of H$\alpha$ for
each XMD galaxy individually (along with available HI maps for some of them),
and then, we summarise our conclusions for the subsample as a whole.

\subsection{UGC~772 = SDSS J0113+0052}

This galaxy has quite a large extent ($\sim1$ arcmin) and very complex
morphology and kinematics. Observations were obtained at rather poor
seeing of $\sim$4 arcsec, but this was sufficient for general
characterisation and modelling of its velocity field. The seeing probably
has somewhat affected the picture for the `S knot',  which has a total extent
of $\sim20\times10$ arcsec.
The optical light, both in broad-band continuum and in H$\alpha$, shows
clearly two distinct components. The larger one (or the main body) with the
overall extent of $\sim50\times20$ arcsec  is elongated
at PA$\sim$60\degr\
and contains three HII regions, each having some
elongation and/or substructure. \citet{IT07} obtained spectra for 3 HII
regions in the main body on one slit. Their values of O/H in all three
regions are consistent to each other within the cited errors and can be
treated as O/H for the whole main body.
We estimated that the weighted mean for those three O/H values
corresponds to 12+$\log$(O/H)=7.29$\pm$0.05.
The general velocity gradient in the main body is along the NS direction,
that is almost perpendicular to the major axis of symmetry.

The centre of the second distinct component (hereafter, UGC~772~S) is
situated $\sim$35 arcsec (or $\sim$2.8~kpc at the adopted distance of
16.3~Mpc) south  of the geometrical centre of the main body.
This is also elongated in both continuum and H$\alpha$, roughly in the NS
direction and has a clear structure. There are no published measurements of
O/H for this region. Our recent spectrum of this region at the SAO 6-m
telescope (Pustilnik et al., in preparation) results in 12+$\log$(O/H)
=7.38$\pm$0.07. Despite this O/H being somewhat larger than that in the main
body, the errors of both values are too large to claim a real difference.
The general velocity gradient along the EW direction (about minor axis) is
clearly seen in this component.

The velocity dispersion is rather low (10-20~\kms) in all HII-regions around
current SF sites, where the intensity of H$\alpha$ show local peaks. The
latter is common for the majority of other studied SF galaxies. It reaches
values of
40-50~\kms\ in several low-brightness small periphery knots, where this
enhancement is probably related to shocks in expanding shells.

To zeroth order, the tilted-ring model is a reasonable approximation
for the global motions in this object over the full range of the observed velocities.
However, since    the distribution of H$\alpha$ emission has a huge gap between the main body and
the S component, it is difficult to find a good fit based only on
our FPI data. The GMRT  HI velocity field for this object from
\citet{Ekta08}, which shows consistent features with our H$\alpha$ velocity
field in the overlapping regions, is helpful for the choice of the best-fitting
model. Therefore we accepted the value of disc inclination from those HI maps.
The next parameters -- the systemic velocity of kinematic centre, and the
$PA$ of major kinematic axis were derived from our H$\alpha$ data.
There is no well defined photometric centre. Therefore, the
rotation centre was determined only from the symmetry of the velocity
field. Its position also agrees (within a few arcsec) with the centre
of HI distribution and rotation as it follows from the GMRT data.
 A model was constructed only for $r > 12$ arcsec; since in the inner
region the number of points with the measured velocities is small and we
fail to construct the stable model. In the outer regions the circular rotation
model fits well the observed velocity field. The values of residual velocities
in general do not exceed  5--10~\kms, reaching $\pm$15~\kms\ in several
regions.
The latter are still smaller than the observed amplitude of
the rotation curve (18~\kms). The largest non-random residual velocities
(both negative and positive) are seen in the region of UGC~772~S. This looks
like a more or less systematic gradient with the full range of $\sim$30~\kms.

Due to the patchy structure of ionized gas and the evident problems with
a simple modelization of its global velocity field, it is important to compare
the H$\alpha$ velocity field with a lower resolution, but much better sampled
motions of HI as seen, e.g., in HI maps obtained with GMRT in \cite{Ekta08}.
In particular, as well seen in their  Fig.~9  (right) for the FWHM beam
of 15$\times$10~arcsec, the gas velocities observed in HI, in both
the main body and UGC~772~S, correspond well to those measured in H$\alpha$.
The general direction of the HI velocity
gradient is close to NS within the main optical body in accord with the
ionized gas motions. However, one can see strong deviations of HI velocities
near UGC~772~S (and the related HI-density peak), and  in the southern
part of the HI `disc'. They look like an additional, smaller  velocity
component with the gradient being roughly along the EW direction. The other
indications
on the strong tidal HI protrusions are seen in their map in  Fig.~9
(left).
The H$\alpha$ velocity field in UGC~772~S also shows a gradient close to the
EW direction, and thus, ionized gas motions are coherent with those of HI.
These kinematically and spatially decoupled components of UGC~772 suggest
that we are witnessing a dwarf galaxy merger stage (the main body of UGC~772
and UGC~772~S),  probably  close to coalescence.
An alternative interpretation of the region UGC~772~S as a supergiant
HII region on the edge of a dwarf galaxy looks quite improbable due to three
facts:  (i) the strong disturbance of the overall HI morphology, (ii) the
occurrence of the galaxy's highest HI density peak in the edge region is very
unusual for an isolated  galaxy, and (iii) the velocity
field in this region is clearly detached from the main galaxy motions.

UGC~772 is situated in a small group of gas-rich galaxies
\citep[e.g.,][]{Ekta08}. Therefore, such galaxy collision looks quite
probable. The current starburst in UGC~772 and in this smaller southern
'companion' galaxy is then most probably triggered by their collision.

\subsection{UGC 993 = HS~0122+0743 }

This is one more rather extended object with very complex morphology and
kinematics. The body, in both continuum and H$\alpha$, represents a group of
seven bright knots (SF regions) superimposed on a LSB component. They are
visually joined into two aggregates of similar size.
The velocity field looks rather regular with two regions of clear gradient
with  the same direction and sign and the velocity ranges of $\sim$100 and
$\sim$60~\kms. The regions roughly correspond to the W and E parts of the
body. The velocity dispersion in most of the galaxy body, and in particular,
around the SF knots, is quite low: $\sim$10-30~\kms. As in the previous
galaxy, it is enhanced up to 50-60~\kms\ in  several areas  between
HII regions.

Guided by its morphology and the appearance of its velocity field, we model
this object with two rotating discs in contact (see
Figure~\ref{fig:results_p1}). The first component, on the western edge,
is fitted with $V_{\rm sys}$=2953$\pm$2~\kms,
$PA$=295\degr$\pm$5\degr, and with model-derived
$i=37\degr\pm4\degr$. Maximal rotation velocities (after inclination
correction) reach $\sim$30~\kms. The second component, at the eastern
edge, is best modelled with $V_{\rm sys}$=2927$\pm$2~\kms,
$PA$=285\degr$\pm$3\degr, and  $i$=69\degr\  (estimated from its axial ratio).
For this component, the inclination corrected, maximal
rotation velocity reaches $\sim$50~\kms. Resulting residual velocities appear
rather small ($\lesssim$8~\kms) in most of the mapped area. The largest
positive values of residuals are seen south of the bright HII region
of the western component and on the NW of the eastern component. They may be
the tracers of current interaction or outflows. The two prominent, but
relatively compact regions of negative residuals (near the kinematic centre
and near the SE edge) in the eastern component can be related to shells
around recent starbursts. Or alternatively, the latter can be the tidally
disturbed gas with counter-motion relative to that at the NW edge.
For the case of modelling UGC~993 by a single rotating disc (as the first
guess), the overall residual velocities were by a factor of two and
more larger.

The recent HI maps obtained at GMRT (Ekta et al. 2009, in preparation) show
clear evidence of disturbed structure, with the large low-density HI plume
stretching to the angular distance of $\sim1$ arcmin to the south of a
denser HI envelope that covers the optical body of the galaxy. The whole HI
gas shows very complex kinematics. In the regions, where the H$\alpha$
emission is well traced, HI displays the velocity pattern similar to that
presented here in H$\alpha$.

All available morphology and kinematics data suggest that in UGC~993, we are
witnessing a rather rare case of dwarf galaxy encounter in contact, when
they have well traced regular velocity fields of both components. The
kinematic data suggests that the real distance between the colliding galaxies
is significantly larger, and their visible contact is due to projection
effects.
The strong SF activity in many HII regions, presumably triggered
by the recent strong interaction, is spread over the components' bodies.

From the model fit of the observed velocity field, the deprojected rotation
velocity in the E component is a factor of $\sim$1.6 larger, and hence, its
estimated total mass should be also significantly larger.
If the Tully-Fisher scaling is valid for the two discs in collision
(M$\varpropto$V$^{3.5}$), the E component can be $\sim$5 times more massive.
This would imply a case of minor merger.
However, rather similar sizes of the two galaxies in question (as best
followed namely through H$\alpha$ velocity field) and a strong HI tidal plum
mentioned above, are strong evidence for a major merger. The above
large difference in rotation velocities can be due to two factors:
(i) not properly
correcting for inclination angle in the W component, (ii) the velocity fields
themselves are
affected by strong interactions, so the Tully-Fisher relation may be not
applicable.
To construct a more self-consistent model,
one needs to combine both H$\alpha$ and HI velocity fields, as well as some
additional photometry including near-IR data. This will be attempted in a
forthcoming analysis.

\subsection{The system SBS~0335--052E,W}

The SF galaxy SBS~0335--052W is the most metal-poor gas-rich object
with an O/H range in different knots between
6.9 $<$ 12+$\log$(O/H) $<$7.22 \citep{Izotov09}.
With SBS~0335--052E \citep[at 22 kpc in projection,][]{PustilnikVLA}
it comprises a unique pair of interacting/merging gas-rich dwarfs with the
lowest metallicities. Both galaxies show unusually blue stellar population
in the outer parts of their optical
images \citep{SBS0335}. The ratio
M(HI)/L$_{\rm B}$$\sim$5 for SBS~0335--052W is one of the greatest known.
The recent detailed study of its disturbed HI kinematics and morphology with
GMRT is presented by \citet{SBS0335_GMRT}. It suggests a merger of extended
gas bodies soon after the first encounter. However,
the denser central regions of both galaxies, with regions of current SF
and surrounding ionized gas, are separated at a significant distance, and can
be thus less susceptible to an external disturbance. Therefore, one can hope
that kinematics of their innermost ionized gas can keep information on galaxy
original rotation.
Due to the recent close encounter between E and W galaxies, the gas motions
are significantly affected by the tidal torques, especially in the outer
parts of colliding galaxies. On the other hand, the gas kinematics
in the innermost parts can be affected by current/recent SF episodes.

One of the important issues in understanding this unusual interacting pair is
the following: why do the optical luminosities and SFRs of E and W galaxies
differ by an order of magnitude, while their global HI properties are pretty
similar? In principle, the substantial part of the answer can be related to
differences in the geometry of collision for two components and the related
strength of tidal action. Indeed, the high resolution GMRT maps of the
innermost HI gas give evidence for a significant difference in the main
velocity  gradients in E and W galaxies.
For the orbital plane sky projection close to the EW direction, the
collision of the E galaxy is closer to a retrograde one. Meanwhile, the
innermost velocity field of the W galaxy,  having its  main gradient in
the N-S direction,  corresponds to an almost polar collision.
This would imply a less effective tidal perturbation. Our FPI H$\alpha$
velocity field allows a consistent check of the HI data.

\subsubsection{SBS~0335--052W}

The observed ionized gas velocity field of the W galaxy looks like a
combination
of two components. One, centred at the bright HII region, shows a very small
gradient, with an amplitude of less than $\sim$10~\kms\ and the direction
close to N--S. The other reflects the gas motion between two peaks
of H$\alpha$ emission (mainly in E--W direction) and has the amplitude of
$\sim$20-25~\kms.
The `best-fitting' model of rotating disc describes primarily the latter motion,
which seems to be related to the tidal flows seen in HI maps. However, the
PA $\sim$70\degr, accounts for the former component. Thus, the H$\alpha$
kinematics gives additional evidence on the presence of a velocity component
roughly in N-S direction, well seen in the high-resolution HI maps.
The fainter HII region, 3 arcsec  east of the main star-forming region,
is situated along the direction, which is closer to the direction of  the
compact HI tidal protrusion, well seen in Fig.~5 of \cite{SBS0335_GMRT}.
By analogy with the discussed opportunity of stimulated SF in asymmetric
shells prompted by tidal outflows in the case of E galaxy
\citep[][]{SBS0335_GMRT},
the similar situation can take place for W galaxy (albeit, with smaller
power).

\subsubsection{SBS~0335--052E}

The ionized gas kinematics of SBS~0335--052E were studied by \citet{GIRAFFE}
on the limited region of $11\times7$ arcsec around the central
star-forming regions. In that work the general velocity field was not discussed. Instead, they
find the two-component H$\alpha$-line profiles in several periphery regions,
with component separation of $\sim$50~\kms, which were treated as the
appearance of fast shells. In our observations we constructed the H$\alpha$
velocity field in the region of $\sim17\times17$ arcsec in size.

Similar to the W component, we compare the motions of ionized gas in the
E galaxy with that of HI gas in the same region.
This comparison is conducted on the GMRT HI maps with the
angular resolutions of $\sim$20 arcsec  and 9 arcsec. Since the respective
analysis was already performed in \citet{SBS0335_GMRT},  this allows us to
look at the observed H$\alpha$ velocities in a more general context.
Namely, there are two systems in the HI velocity field, which are
identified with gas motions in
elongated tidal tails. A longer and more rarefied one is directed in
approximately SW-NE direction with velocities growing to the NE edge, that is
in direction opposite to the position of the companion. A shorter, curved
denser
tail starts roughly in direction to NW and then bends to the W companion.
For this tail (protrusion), the velocity decreases while gas travels further
from the E galaxy. This direction is close to that of the asymmetric arc of
ionized gas well seen in the \textit{HST} $V$-band image of \citet{Thuan97}.

\begin{figure*}
\centerline{
\includegraphics[width=4.1 cm]{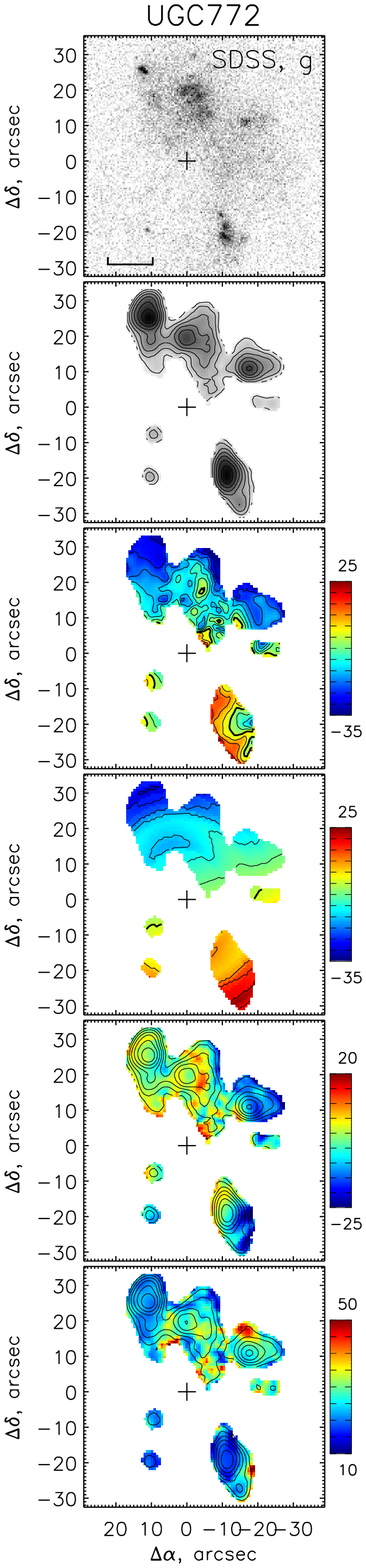}
\includegraphics[width=4.1 cm]{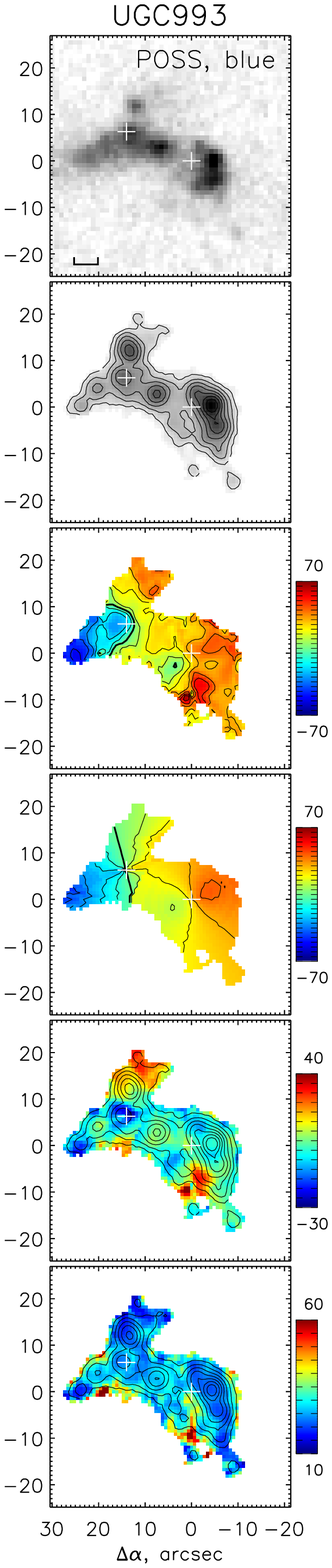}
\includegraphics[width=4.1 cm]{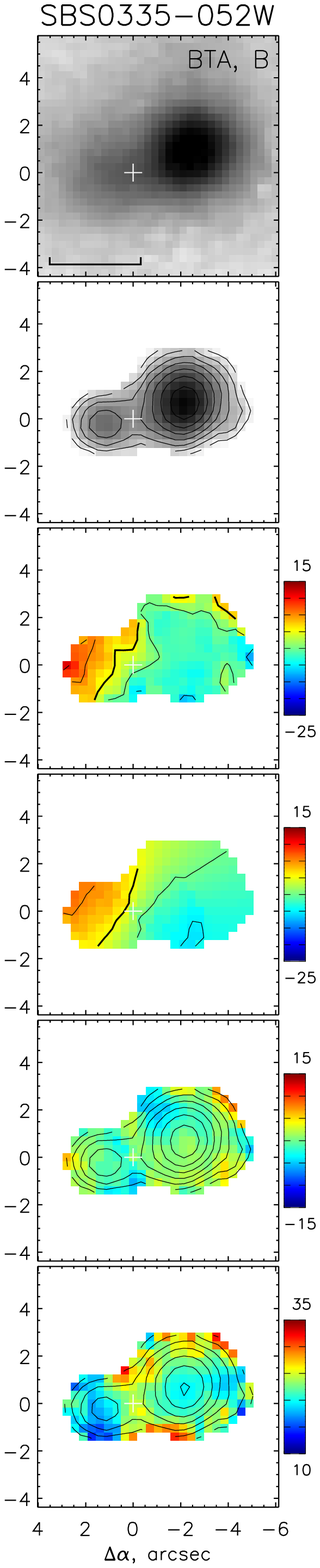}
\includegraphics[width=4.1 cm]{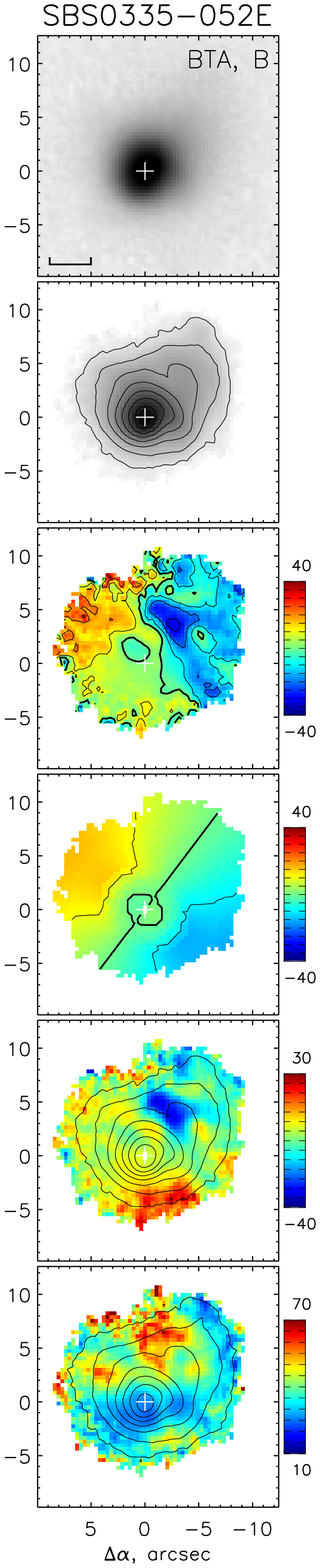}
}
\caption{Images and FPI maps for four program galaxies UGC~772, UGC~993,
SBS~0335--052W and SBS~0335--052E.
From top to bottom: broad-band and $H\alpha$ images in logarithmic intensity
scale; the line-of-sight velocity field in H$\alpha$ and model of circular-rotated disc,
the scale is in \kms\ relative to the systemic velocity which is  corresponded to thick contour;
the maps of residual velocities and of velocity dispersion. In two latter
maps the isophotes of H$\alpha$-line intensity are also superimposed. The
cross marks the kinematic centre. For UGC~993 and HS~2236+1344, centres of
both discussed components are marked. The step between isovelocities is
5~\kms, with the exception of UGC~993, SBS~0335--052E and Anon J012544+075957 where  the step of 10~\kms\ was used. The scale bar in the top panel corresponds to 1 kpc.}
	\label{fig:results_p1}
\end{figure*}

\begin{figure*}
\centerline{
\includegraphics[width=4.1 cm]{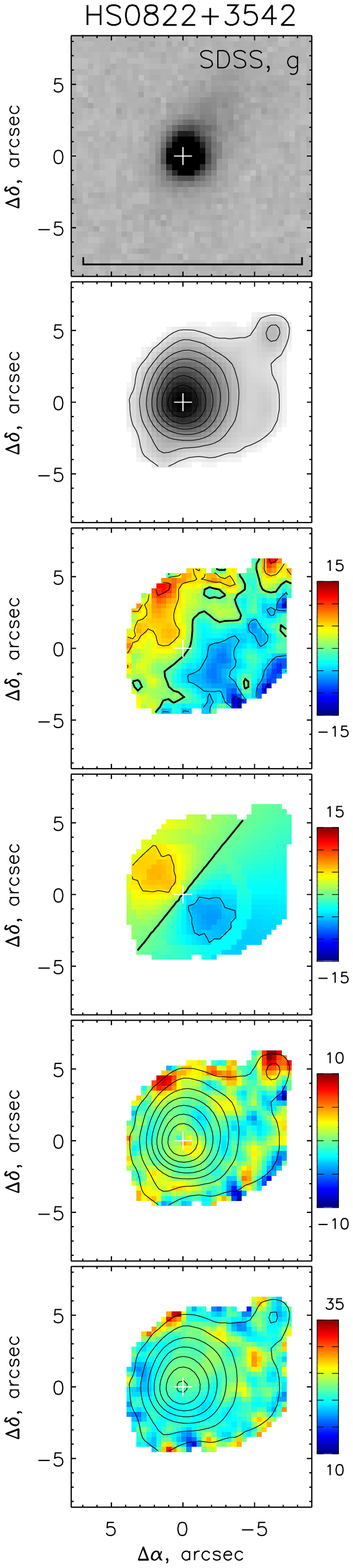}
\includegraphics[width=4.1 cm]{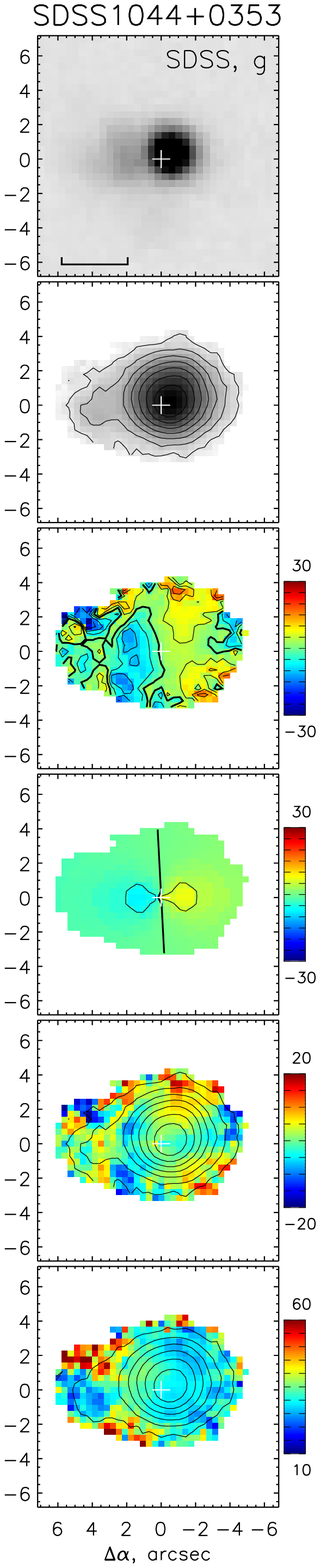}
\includegraphics[width=4.1 cm]{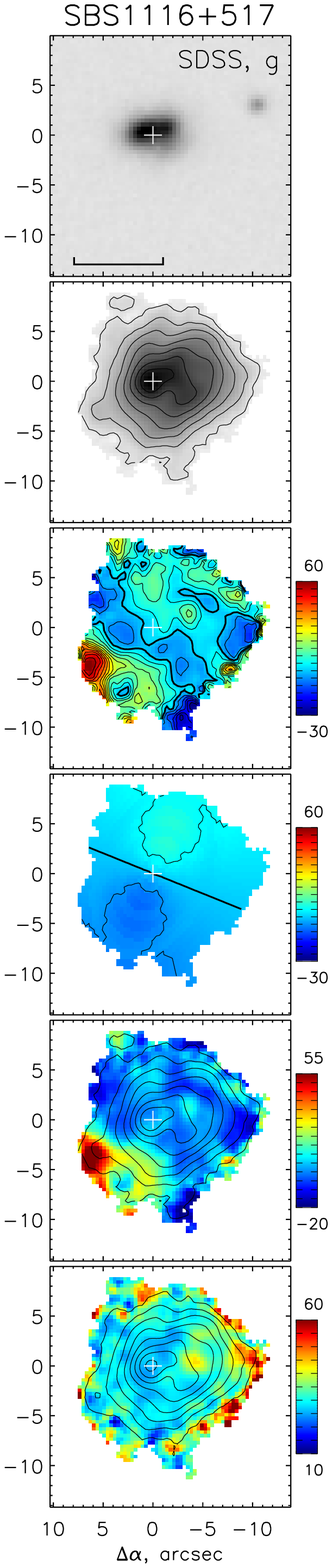}
\includegraphics[width=4.1 cm]{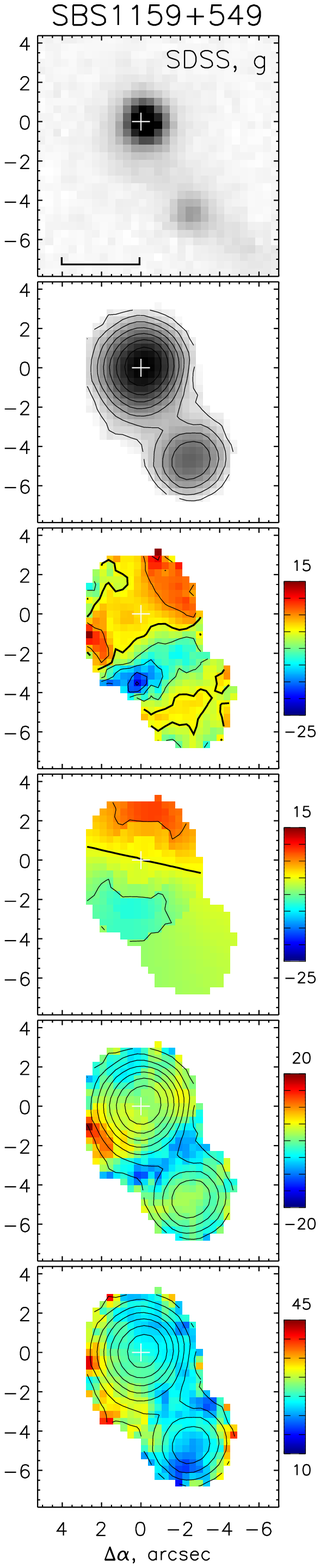}
}
\caption{Same as in Fig.~\ref{fig:results_p1}, but for galaxies
HS~0822+3542, SDSS J1044+0353, SBS~1116+517 and SBS~1159+549. }
	\label{fig:results_p2}
\end{figure*}

The coarse comparison of the observed H$\alpha$ velocity field in Fig.
\ref{fig:results_p1} with that of the described HI one, shows qualitative
agreement.
The region of minimal line-of-sight velocities at $\sim$6--7~arcsec
NW from the central starburst is close to the position of the arc. Thus, the effect of the expanding shell
(visible as relative negative velocities on its front side)  can
also contribute to the structure of the velocity field.
The significant swing of the isovelocity contours in the central
region ($r\leq10$ arcsec  from the centre) seems may be related to the
appearance of an expanding shell. To check how the disturbed velocity field
looks like in the presence of a shell (in the zeroth approximation)
we consider in the Appendix~\ref{App1} several simulated velocity fields
of rotating disc with a shell in different positions. The observed velocity
field of SBS~0335--052E resembles that in our Model~3, which implies an
off-centre location of the shell centre on the disc minor axis.
Based on conclusions, obtained in Appendix~\ref{App1}, we fixed the position
angle of the model disc ($PA=53$\degr) in accord with the orientation of the
large-scale HI structure.
This model of rotation takes into account the main component of the velocity
gradient (roughly from SW to NE). In the residual map one can clearly
separate the same velocity minimum of the irregular form NW from the centre,
related to the arc. The latter is natural to attribute to the approaching
part of a shell, produced during a relatively recent episode of SF. Having
in mind its irregular form, one can think of a superposition of two spatially
close shells. Furthermore, the negative gradient in the NW direction is still
visible, consistent with the similar flow seen along the HI  NW tail.

It is worth noting the unusual behaviour in the model fit of rotation
velocity in the very centre of the galaxy, at $r<2$ arcsec.
This can be treated as an appearance of kinematically decoupled
counter-rotation, which could be a trace of merging gaseous clouds
with the specific orientations of their angular momenta.
However, the observed rotation velocities are too small. Therefore, the more
probable option is that this a feature of the velocity field appears due to
non-circular motions produced by a shell formed near the disc centre.

The galaxy SBS 0335--052E, unlike SBS 0335--052W, shows much more
active star formation during the last several hundred Myr. This induces both
large-scale shells/arcs and increases the amplitude of chaotic (turbulent)
motions. The local peculiarities of H$\alpha$ kinematics are better acquired
by high-resolution spectra with VLT in \citet{GIRAFFE}.
In particular, they detected wide-spread regions outside the central
brightest clusters 1 and 2, where the line profiles are split on two
components with velocity differences of $\sim$50~\kms. The `broad' lines
with FWHM up to $\sim$100~\kms\ are present in these regions as well.
While our angular resolution is about a factor
of 1.7 worse, we also see in the corresponding regions
the enhanced velocity dispersion of $\sim$60--70~\kms, corresponding to FWHM
$\sim$150~\kms. Accounting for smoothing due to the worse seeing, these
values look consistent. In fact, as the illustration in Fig.~\ref{fig:line_prof} shows, our data also hint at the two-peak structure
of line profiles in the mentioned regions, but due to insufficient velocity
resolution they can be treated only as an indication.

Summarising the overall ionized gas kinematics of the E galaxy, we conclude,
that in accord with HI kinematics \citep{SBS0335_GMRT}, its direction of
rotation is closer to the orbital plane. In general, we assume that
before the merging galaxies experience coalescence, their innermost regions
keep the information on their original rotation. In difference with
the W galaxy case (polar collision), the collision of the E galaxy is closer
to prograde or retrograde.
As emphasized by \citet{SBS0335_GMRT},  the latter may be one of the
reasons
of large differences in the current and earlier SFRs between E and W galaxies.

\subsection{HS~0822+3542}

The H$\alpha$ velocity field for this BCG is rather well fitted by the
model of flat rotating disc. The rotation
velocity curve raises up to $\sim$13~\kms\  at the radius $r =3$ arcsec
($\sim$200 pc) and then falls to $\sim7-9$~\kms\  at radius of
$r \sim5-6$ arcsec. The region with the maximal residual velocity (of
$\sim$10~\kms) is positioned at a secondary peak of H$\alpha$ emission at
$\sim7-8$ arcsec NW of  the main starburst region. This could be the
brightest fragment of a large loop, a probable relic
of a somewhat earlier star-forming episode.
The comparison of H$\alpha$ kinematics with HI morphology and
kinematics \citep[see][]{Chengalur06} leads to interesting conclusions.
At the highest angular resolution of $\sim6$ arcsec, the related HI cloud
is strongly elongated in the N--S direction. The brightest HII region is
situated between two HI density peaks. Due to low S/N ratio, the HI
velocity field is constructed only for the low-resolution datacube
(beam$\sim$25 arcsec). This has a general (but small) gradient in
approximately
N--S direction. However, in the vicinity of the optical body one can see the
swing of isovelocity line, roughly in the direction close to the PA of
H$\alpha$ velocity gradient.

The most probable reason for the formation of decoupled kinematic
patterns - the internal as visible in ionized gas, and near polar orbits  as
outlined by HI - is the interaction with a nearby LSB dwarf SAO 0822+3545.
The latter is also mapped in H$\alpha$ in the same observations and is
discussed in a Subsection~\ref{sec:lsb}. This LSBD is situated at the projected
distance of 3.8 arcmin ($\sim$15~kpc) and has line-of-sight velocity of only
$\sim$15~\kms\ larger. It is $\sim$3 times more massive in HI and
$\sim$0.35~mag more luminous in $B$-band than HS~0822+3542
\citep{SAO0822,Chengalur06}. Therefore, one expects that the past
interaction of the BCG progenitor with this LSBD could induce a starburst in
HS 0822+3542.

\citet{Corbin05}, from their  Hubble Space Telescope imaging of
HS~0822+3542 with the spatial resolution of $\sim$5~pc, found that
its central SF region consists of two distinct components with projected
separation of $\sim$80~pc. Based on this morphology, they suggest that
this very compact dwarf galaxy is currently being assembled from two tiny
unrelated components.
Our H$\alpha$ kinematics  of this BCG, in difference to several other
galaxies in this study, is rather well fit by a single disc [albeit
disturbed by the earlier starburst(s) and related shell(s)]. This is
difficult to match with the idea that the BCG is currently forming from
two unrelated fragments.

It is worth noting that this H$\alpha$ kinematics and HI data from
\citet{Chengalur06} for HS~0822+3542 do not provide a comprehensive picture
of the overall
gas motions. Indeed, on one hand, the regular H$\alpha$ rotation within the
whole region of optical emission (with the total extent of $\sim$16 arcsec, or
$\sim$1~kpc), has an amplitude of $\sim$13~\kms.  On the other hand, HI
data shows rotation with an amplitude of less than 5~\kms.  At the moment
it is unclear how much the beam-smearing affects the HI velocity data.

\subsection{SDSS J1044+0353}

In the SDSS colour image, this galaxy is elongated along the E--W direction,
with an extent of $\sim$10 arcsec. The extent along the N--S direction is
$\sim$4 arcsec. The brightest part,  the `comet head' on the W edge, is very
blue; while the more diffuse tail with a second, eastern `peak' at
$\sim$4 arcsec is somewhat redder. The analysis of surface brightness and
colour radial dependencies by \citet{Papa08} confirms this impression.
They use the blue colours of a lower SB part of this object to infer the
small ages of the underlying LSB disc. However, reservation are made on
the possible significant contribution of nebular emission to its colour,
that should be accounted for. Indeed, from our rather deep H$\alpha$ image,
the extent of the galaxy nebular emission ($\sim12\times8$ arcsec)
appears to be similar to that of the SDSS continuum images.

The two-nuclei morphology of this galaxy is similar to that of
other studied objects (HS~2236+1344,  SBS~1159+545),
for which their H$\alpha$ kinematics show good evidence for a merger event.
The simple disc model that best fits the observed H$\alpha$ velocity field has
the following parameters:
$V_{\rm sys}$ = 3865$\pm$1~\kms, $PA$=273\degr, $i$=51\degr.
The bright rotating core is well fit by a rotation centre shifted to
the east by $\sim$0.5 arcsec from the photometric centre. The maximal
observed (not corrected for inclination) rotation velocity is 8--9~\kms.
This peak is well seen on the model rotation curve in Fig.~\ref{fig:models}
at $r \sim2$ arcsec. The residual velocities in the `core'
region ($r < 2.5$ arcsec) are small ($<$ 5~\kms). In outer regions the
regular rotation falls significantly,
and the residual velocities amount up to 10--15~\kms.
The fitting by a flat disc model with the maximal rotation
velocity of $\sim$10~\kms\ results in rather small overall residuals.
However, the two outer regions of the E component show residuals of up to
$\pm$20~\kms. Namely in these regions we see a maximal velocity dispersion
of up to 50~\kms.
Besides,  the radial dependence of rotation velocity is quite atypical
with a substantial fall after the maximum at $r \sim2$ arcsec. This
suggests that the gas kinematics
of the eastern half of the galaxy are significantly disturbed. On the colour
SDSS image, one can also see a redder, diffuse extension south of the main
elongated body. All these properties are evidence for an unrelaxed state of
this galaxy.

The available data are not sufficient to make a confident conclusion on
the nature of this object. At the current level of understanding of its
morphology and gas kinematics,  the data do not contradict an idea that
this is the result of a relatively minor, almost completed merger.
Other interpretations look more problematical.
The eastern component has a significantly smaller mass  than the western
one, in difference with, e.g., the case of HS~2236+1344. This results in
a relatively weak velocity disturbance in the W component
and a much fainter luminosity of E component in respect of that for W one.
High-resolution HI mapping could probably produce a more complete picture of
its dynamics during the last few hundred Myr.

\subsection{SBS~1116+517 = Arp's galaxy}

This XMD BCG has a complex morphology in the central part, which consists
of three knots with very different fluxes situated in a region with
a total extent of $\sim5$ arcsec \citep[as first noticed by][]{Arp}.
The outer parts also show rather irregular
morphology with finger-like protrusions, in particular on the S and E edges.
While the full range of line-of-sight velocities in H$\alpha$ is more than
80~\kms, the velocity field is rather complex.
It is hard to see tracers of regular rotation.
Nevertheless, we try to construct the model of rotating disc for
radial distances $r <9$ arcsec, which we extrapolate to the outer region.
The centre of the inner isophotes was set as the centre of rotation.
As one expects, the rotation is indeed small:
its maximal observed line-of-sight velocities are of order 5--10~\kms,
while the residual velocities are $\sim$20~\kms\
in the central part and reach the value of $\sim$50~\kms\ at the E and SE
periphery. The region with the most peculiar (redshifted) velocities
is situated South-East from the bright `central' knot and possibly represents
the result of a merger (or infall to the disc).

While the `best-fitting' model with rotation can account for maximal rotation
velocity of $\sim$10--12~\kms, the resulting residual velocity field
shows the regions with values down to --20~\kms\ and up to +50~\kms.
This very disturbed velocity field, especially in outer parts of an ionized
gas body, suggests strong perturbation. The presence of expanding shells from
recent starbursts
is probably one of the possible reasons. If one assigns the regions
with the most negative velocities on the residual map to approaching fronts
of such shells, their expansion velocities are $\sim$20~\kms\ and the total
sizes are $\sim$3--5~arcsec (or $\sim$300--500~pc at the adopted
distance of 23~Mpc). The sizes are quite
typical of ionized-gas shells, visible in similar galaxies.
However, the large positive residual velocity at the eastern edge of
the galaxy is difficult to reconcile with shells.
The morphology of H$\alpha$-emission region and velocity fields,
both the observed and residual,  suggests strong external
disturbance. No candidate galaxy capable of producing that strong of a
disturbance
is visible in the BCG environment. Therefore, the most natural interpretation
that emerges from the study of SBS~1116+517 morphology and gas kinematics
is the recent major merger of dwarf galaxies, since for a minor merger one
expects a relatively weak effect on the overall velocity field of the more
massive component. If this picture is correct, one should expect that the
HI morphology and velocity field are highly disturbed.
Thus, this galaxy requires follow-up HI mapping with angular resolution of
3--5~arcsec.

\subsection{SBS~1159+545}

On the SDSS image SBS~1159+545 consists of two very blue knots
(separated by $\sim5$ arcsec, or $\sim$1.3~kpc at the adopted distance of
52.5~Mpc), elongated roughly along NE--SW direction, with the luminosity
difference of a factor $\sim$2. There are also redder lower brightness
fragments; one is situated between the two blue knots and the other, with
the total extent of $\sim5$ arcsec, stretches as a tail of the fainter,
blue SW knot.
The SW knot itself has an extended structure, roughly oriented along E--W, and
its redder tail looks like a continuation of this blue knot. On the other
hand, the brighter NE knot also looks elongated on our narrow red continuum
image, in a direction of S--N.  This overall complex, curved morphology indicates
a non-equilibrium state of this object.

The full range of H$\alpha$ velocities seen in our data is $\sim$40~\kms.
The velocity field is complex, and does not resemble a projected rotation.
Moreover, it resembles that of some other objects in this study, with
`two-nuclei' morphology, which we have argued are mergers
close to the full coalescence (SDSS J1044+0353 and HS~2236+1344).
Namely, there are two local maxima in the line-of-sight velocity, close
to the positions of the H$\alpha$-emission peaks and a clear minimum along
the border, or `middle' lane, between these SF knots.
From the position of this middle lane, the observed velocities grow to both
NE and SW directions. However, after reaching the local maxima near the
H$\alpha$ emission peaks  they again fall. The largest velocities are
seen on the W and E edges of the NE knot as well as at similar outer parts
of the SW knot.

The velocity field model was built for the range $r < 5$ arcsec (with
the centre at the NE knot) and was extrapolated to the SW knot.
The `best-fitting' model for the flat rotating disc centred on the
brighter NE knot can account for only an amplitude of $\sim$8~\kms\
along the line-of-sight.
The residual velocities are significantly larger, amounting on the knot's east
edge of 15--20~\kms. Similar indications of an overdisturbed velocity field
come from the map of velocity dispersion. While in the regions around bright
knots, this
parameter does not exceed 15--20~\kms. In the E part of NE knot and the
W part of the SW knot, this exceeds 30--50 \kms. The
$PA$=--13\degr\
of the `best-fitting' kinematic model significantly differs from
$PA$$\approx30$\degr\ of the apparent `disc' plane -- if the two knots were
HII regions in the same disc galaxy. But this is consistent with the
orientation of the NE knot itself, as seen from our narrow band continuum
image. An attempt to interpret the whole
velocity field as a result of combination of two expanding shells does not
work. In this model, it is difficult to understand how the gas in the
middle lane acquired its low observed velocity, while the
line-of-sight velocities in the directions of expected centres of
shells (NE and SW knots)
show local maxima. Summarising this velocity field  analysis, we suggest that
from both the BCG morphology and gas velocity field, the most plausible
interpretation of the data is the merger of two dwarf galaxies.

\subsection{HS~2236+1344}

\begin{figure*}
\centerline{
\includegraphics[width=4.1 cm]{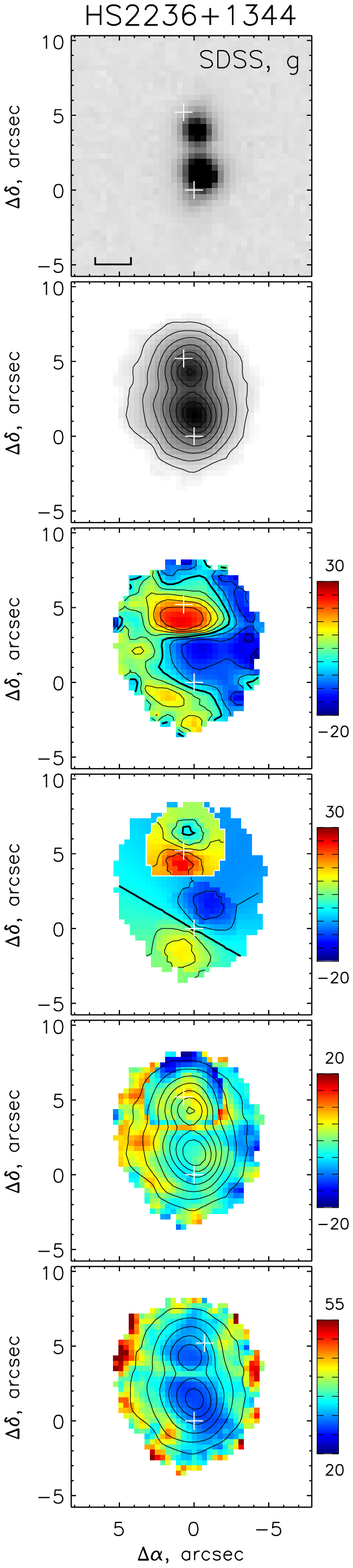}
\includegraphics[width=4.1 cm]{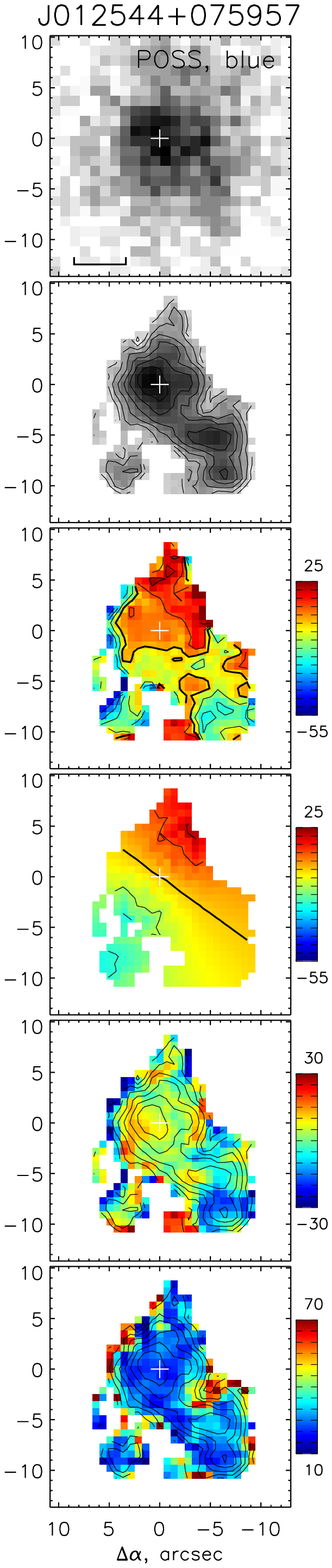}
\includegraphics[width=4.1 cm]{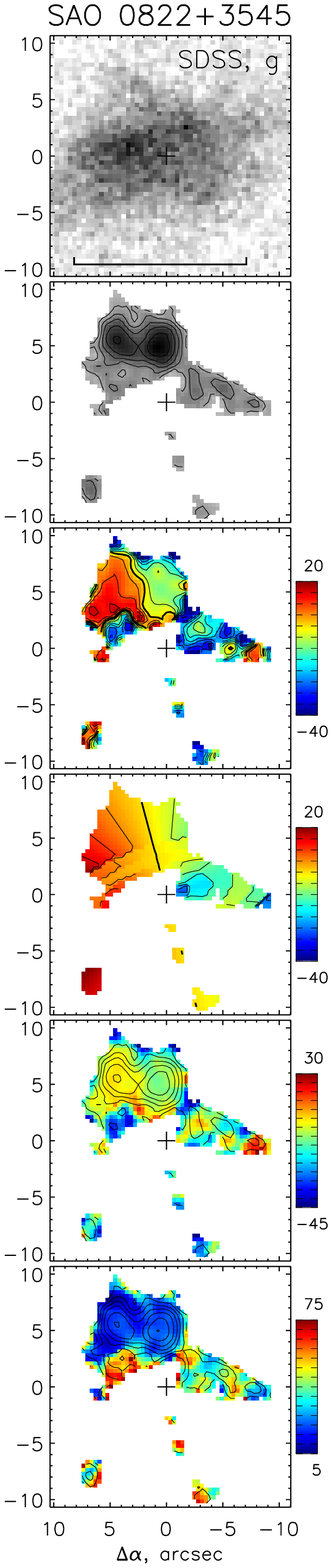}
}
\caption{Same as in Fig.~\ref{fig:results_p1}, but for XMD galaxy
HS~2236+1344 and for two by-product LSB galaxies Anon~J012544+075957
and SAO~0822+3545.}
	\label{fig:results_p3}
\end{figure*}

This XMD BCG also has the two-nuclei morphology with disturbed outer parts
(Fig.~\ref{fig:results_p3}).
The separation of two bright knots, oriented in the N--S direction,
is $\sim3$ arcsec (or $\sim$1.2 kpc in projection at the adopted distance
of 85~Mpc).
As seen in deep images  (not shown here), the latter includes small tails
protruding from both the northern edge of the N component (to the W) and
from the southern part of the S component (Pustilnik et al., in preparation).
These features already hint on the tidal perturbations in both
components. \citet{Pramskij03} studied the H$\alpha$ velocity field of this
object with a long slit, positioned along the N--S direction. They discovered
in the position-velocity diagram that the line-of-sight velocity makes a jump
in the
middle between the two knots. Our   H$\alpha$ velocity field confirms
that finding and adds important new features, including high velocity
dispersion and large residual velocities near the regions of current
starbursts for any model of single rotating disc.

The total  velocity range is $\sim$45~\kms. We attempted to fit the whole
observed velocity field with the model of single flat rotating disc.
Two variants of the circular
motion model were constructed: (i) rotation centre coincides with the brighter
(Southern) knot; (ii) rotation centre is fixed in the middle between Northern
and Southern knots.
However, both variants fail to reproduce the observed velocity gradients.
For an alternative approach, the
approximation of the overall H$\alpha$ velocity  field  by two rotating discs,
positioned near the centres of S and N knots, gives a good fit with maximal
rotation velocities of $\sim$25~\kms\ in both components (see
Table \ref{t:fit_disc} and Fig. \ref{fig:results_p3} and \ref{fig:models}).
These discs have close $PA$, V$_{\rm sys}$, $i$ and maximal V$_{\rm rot}$.

Due to lack of space,  we focus on the model with two discs which provides
a better fit to the observations than either single disc model.
The map of velocity dispersion shows this parameter to be rather large in this BCG.
Even its minimal values, located in the regions of both bright knots, are
of $\sim$25~\kms. In the adjacent regions, this parameter raises till
35-40~\kms, while in the outer parts, the velocity dispersion exceeds 50~\kms.
The observed kinematics can not be explained as well by giant ionized shells
in a single rotating disc.
The latter are commonly expected to show negative relative velocities
in their front edges, that is near the centres of star-forming regions,
while the observed residuals are positive.

Taking into account all available morphological and kinematics data on this
BCG, we conclude that the most likely nature of this object is a close
strongly interacting pair  of dwarf galaxies with comparable masses.
It is quite probable that this pair is in the stage of merger close
to full coalescence. This also
provides a natural interpretation of the starbursts in both
components. Indeed, as summarised by \citet{Begum08}, the late-type
quiet dwarfs with that low V$_{\rm rot}$ have typical M$_{\rm B}$ in the
range of --12.5 to --14.9. Therefore, the M$_{\rm B}$=--16.9 for
HS~2236+1344 indicates a very strong starburst.

\subsection{Properties of two by-product LSB galaxies}

\label{sec:lsb}

In this section we briefly discuss two galaxies that appeared sufficiently
close on spatial and velocity separation to the main studied XMD galaxy
and were mapped along with the main target.

\subsubsection{Anon J012544+075957}

\begin{figure*}
\centerline{
\includegraphics[width=5 cm]{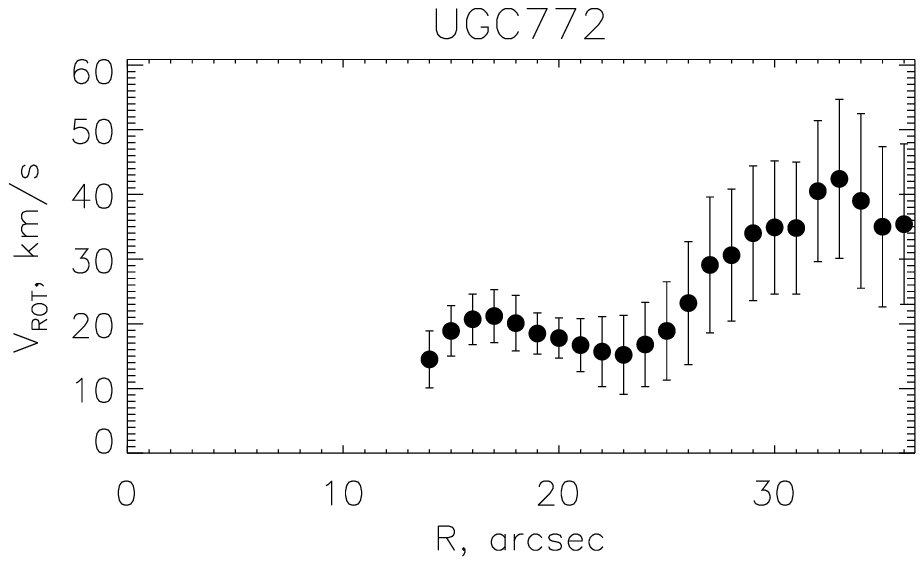}
\includegraphics[width=5 cm]{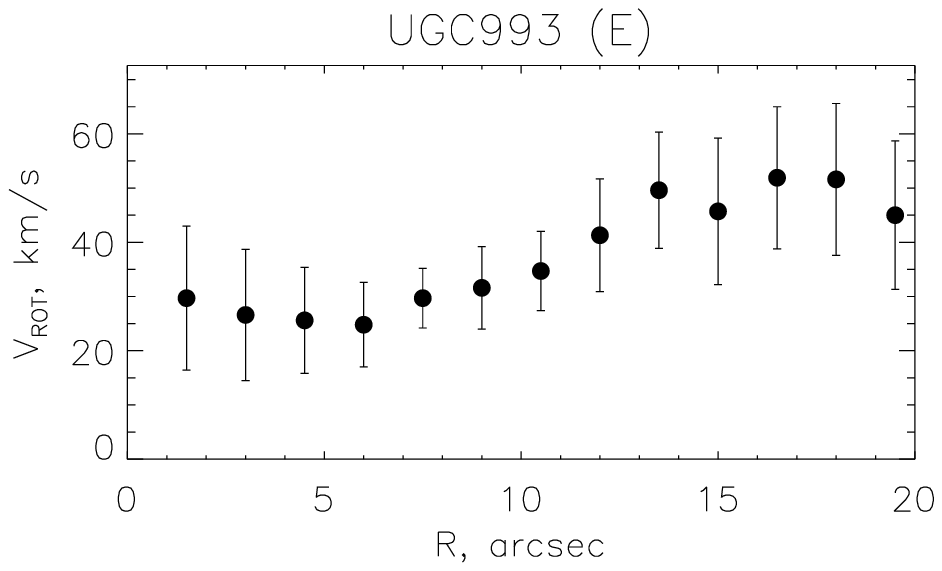}
\includegraphics[width=5 cm]{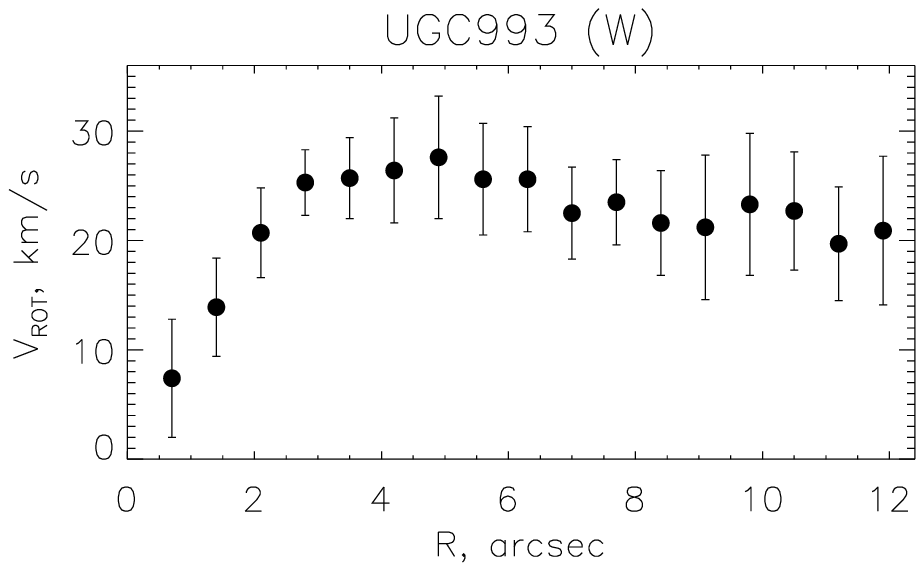}
}
\centerline{
\includegraphics[width=5 cm]{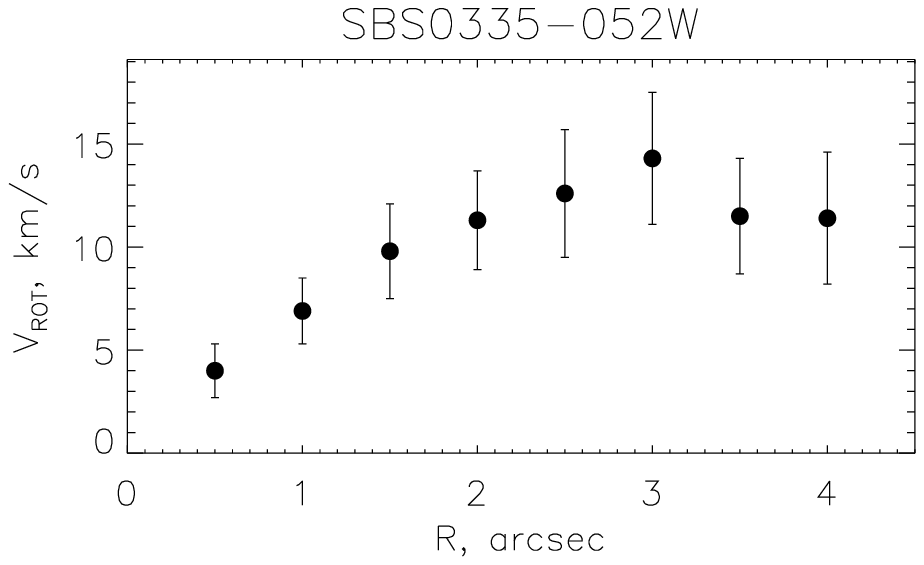}
\includegraphics[width=5 cm]{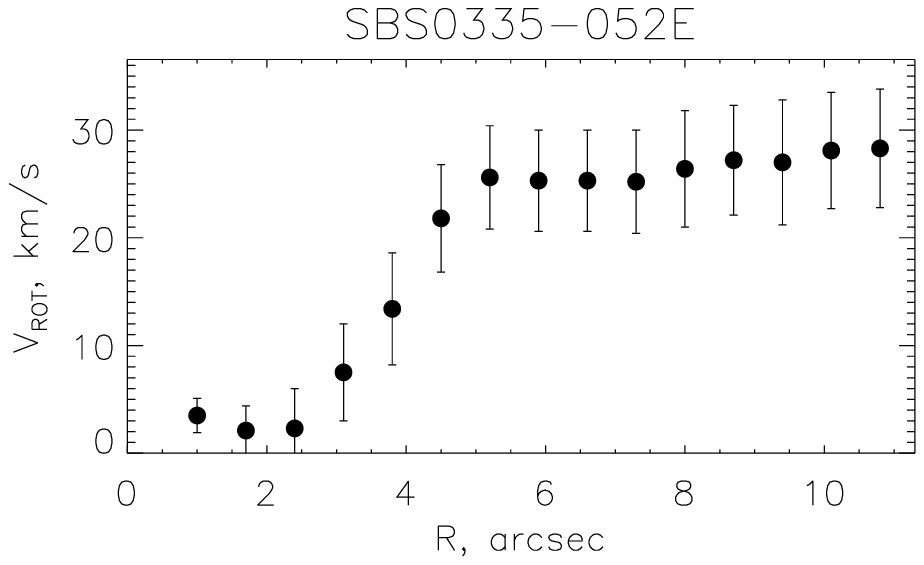}
\includegraphics[width=5 cm]{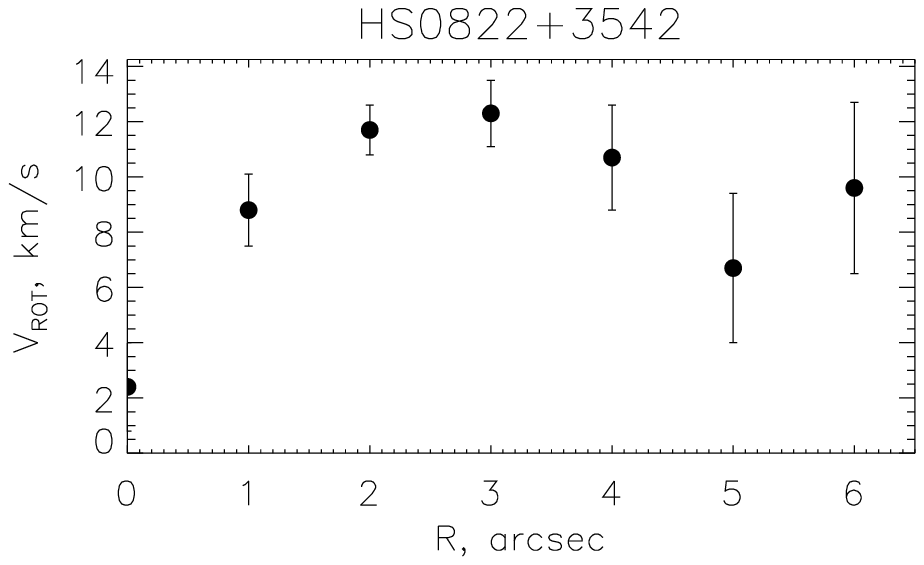}
}
\centerline{
\includegraphics[width=5 cm]{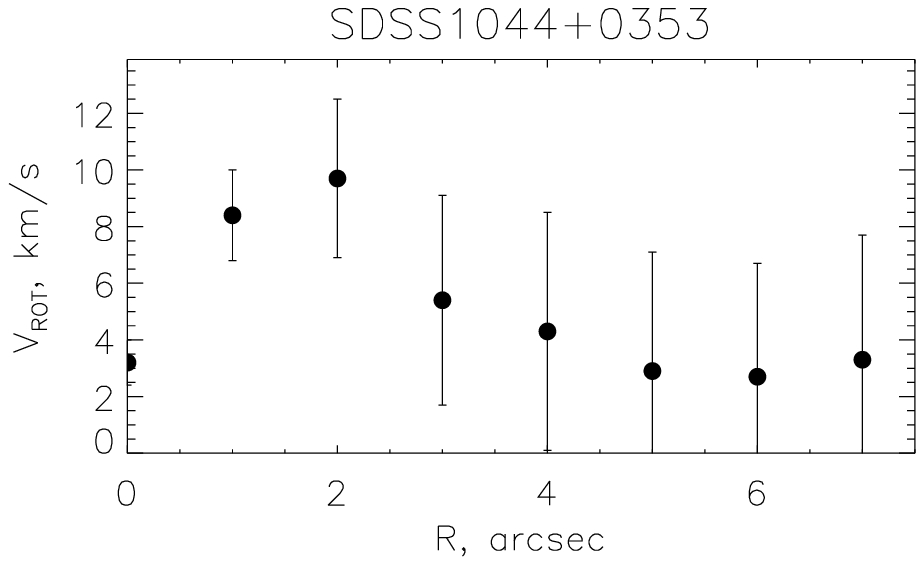}
\includegraphics[width=5 cm]{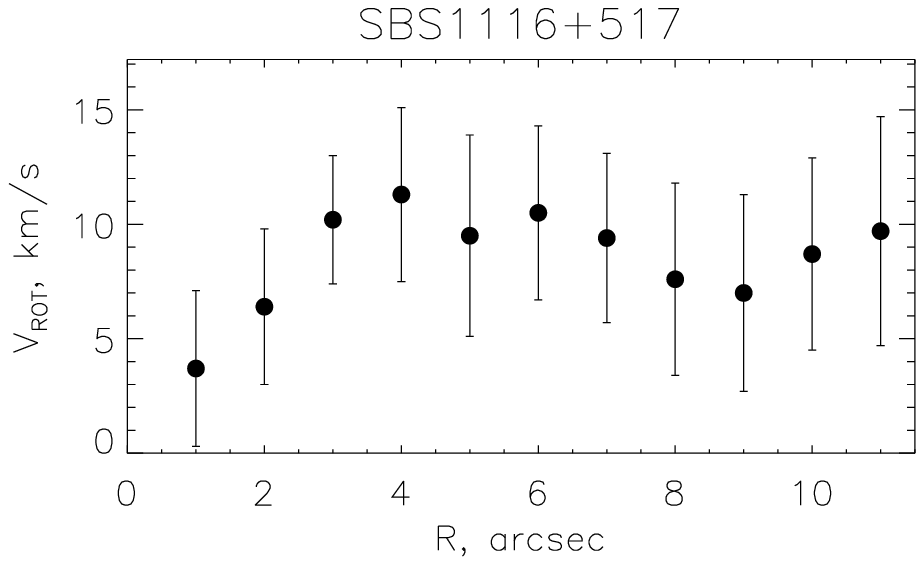}
\includegraphics[width=5 cm]{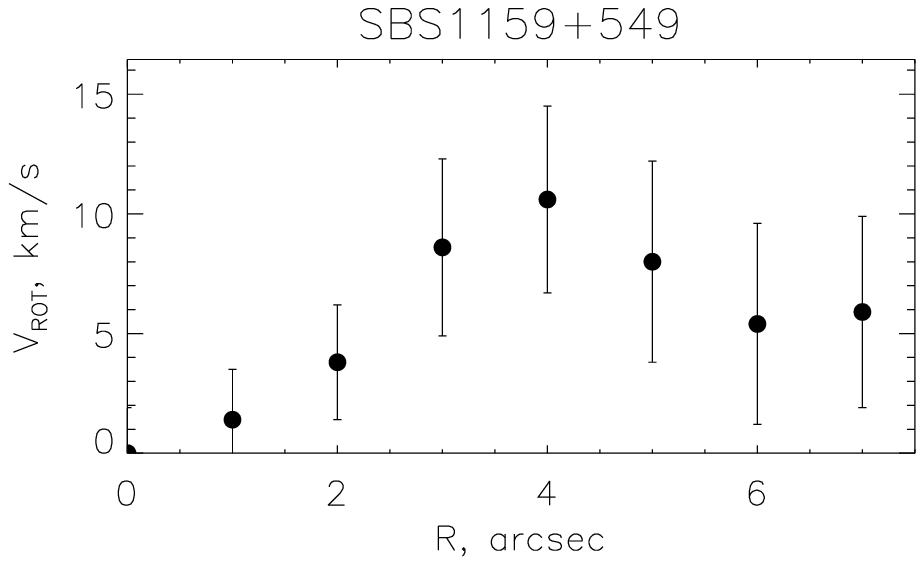}
}
\centerline{
\includegraphics[width=5 cm]{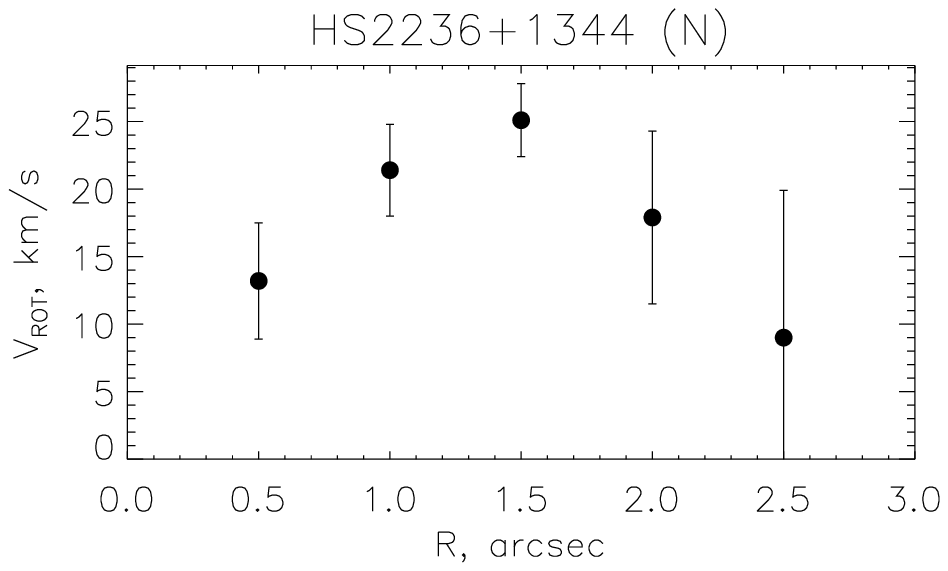}
\includegraphics[width=5 cm]{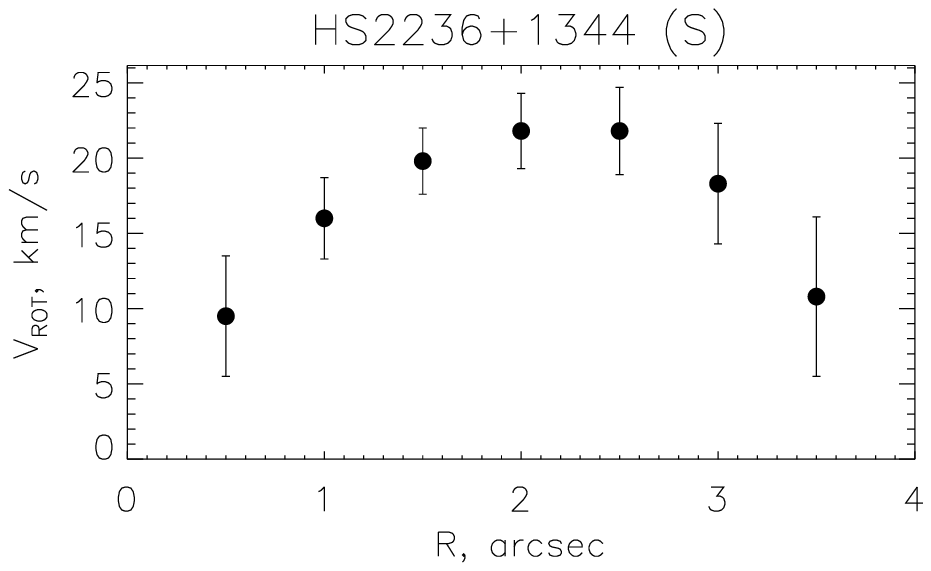}
\includegraphics[width=5 cm]{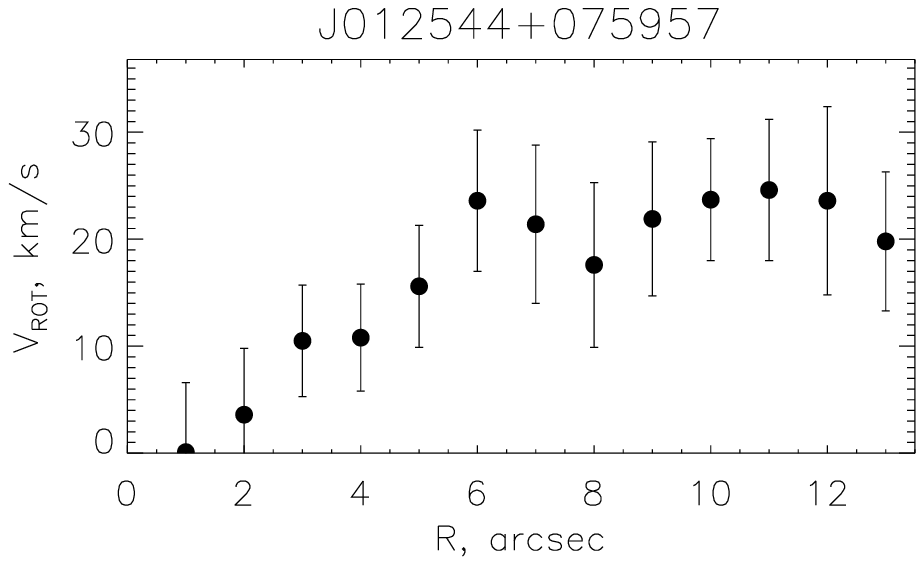}
}
\caption{Radial dependence of rotation velocity in
the best-fitting  models for all discussed objects.
}
	\label{fig:models}
\end{figure*}

In the FPI datacube for UGC~993, an LSB dwarf galaxy Anon J012544+075957
appeared (see bottom of Table~\ref{t:list}) at 2.5 arcmin East and
$\sim40$ arcsec North from the main target ($\sim$29 kpc in projection).
Its line-of-sight velocity (as found here from its H$\alpha$ velocity field)
is $\sim$70~\kms\ larger than the mean velocity of the UGC~993 system.
Its H$\alpha$ image (Fig.~\ref{fig:results_p3})
displays three main regions elongated roughly from NE to
SW, forming  the main body with the total extent of $\sim25$ arcsec
and a separate faint knot SW from this. The velocity range along the body
is $\sim$70~\kms.
The `best-fitting' model  parameters for a rotating thin disc are as follows:
V$_{\rm sys}$=3012~\kms, $PA$=324\degr, $i$=64\degr.
The resulting rotation curve reaches a maximum value of
$\sim$25~\kms\
(Fig.~\ref{fig:models}) at the edge of the H$\alpha$ map ($r \sim12$ arcsec).
The gas morphology of this LSB galaxy is clearly disturbed and
elongated nearly perpendicular to the plane of rotation. It is interesting
that in the blue continuum, the galaxy looks less elongated than in
H$\alpha$. However, there are two extensions approximately to the N and S,
close to the elongation of H$\alpha$ emission.
The velocity dispersion near the peaks of H$\alpha$ emission is a modest
$\sim$25~\kms. In some of the outer regions, it raises up to
60-70~\kms, but this is probably attributed to a low S/N ratio.

Along with the already discussed two kinematically decoupled components of
UGC~993, this dwarf Irr galaxy likely comprises a progenitor triplet of
low-mass galaxies, whose common dynamical evolution resulted in a recent
merger, as seen in UGC~993.
The general problem of such dwarf bound systems was addressed, in
particular, in  \citet{Tully06}.

\subsubsection{SAO 0822+3545}

Another by-product object, SAO~0822+3545, the LSB dwarf galaxy,
known as  a companion of HS~0822+3542, was already discussed in
\citet{SAO0822}. Its unusually blue BVR colours imply a rather small age for
its main stellar population.
Low level SF in this LSBD was noted in the latter study based on the
detection of a low EW H$\alpha$ emission in the long-slit spectrum taken
along the major axis.
The new data show the 2D distribution of H$\alpha$ emission (see
Fig.~\ref{fig:results_p3}) and allow us to perform a more careful analysis of
SF issues.

In particular, our H$\alpha$ image shows that the current SF in this LSB
dwarf is highly inhomogeneous. It is concentrated in two close HII regions
No.~1 and 2
($\sim4$ arcsec, or $\sim$0.26~kpc in between) at the Northern edge of the
main galaxy body, where no light concentration is seen in the broad-band
continuum images. A fainter H$\alpha$ protrusion is seen in the W
part of the faint irregular structure. The H$\alpha$ kinematics are complex.
On one hand, due to very limited spatial coverage of H$\alpha$ emission
over the LSBD body,
it is difficult to derive the parameters  of a rotating disc model
based only on H$\alpha$ data.
On the other hand, the H$\alpha$ velocities
are in good agreement in the overlapping regions with those
observed in HI as presented in \citet{Chengalur06}.
Based on this fact,  we fixed the position of the rotation centre,
$PA$, and $i$, as they emerge from the mentioned HI maps.

As usual, the velocity dispersion near the brightest H$\alpha$ knots is
rather low: $\sim$5-15~\kms.
In several regions outside these two knots, where
H$\alpha$ is still seen - at $\sim4$ arcsec SE of knot No.~1 and in the W
protrusion - the velocity dispersion is large, up to 60-70~\kms. This could be
evidence of recent strong disturbances of the galaxy's ISM.
We also notice that the total H$\alpha$ flux in this LSB galaxy is $\sim$25
times smaller than that in its companion BCG HS~0822+3542, while its total
blue luminosity is $\sim$30 per cent higher \citep{SAO0822}. This seemingly
reflects the difference between the SF properties in the two interacting
galaxies,
related to the difference in their density distribution and the strength of
the tidal trigger.

\section{Summary}
\label{sec:summ}

Summarising the results and discussion above, we draw the following
conclusions:

\begin{enumerate}
\item
The ionized gas kinematics in very low-metallicity star-forming galaxies,
 studied with FPI H$\alpha$ observations are
significantly disturbed and rarely can be well fitted
by only  single  disc with regular rotation.
\item
Several of our galaxies show more or less clear evidence from
morphology and H$\alpha$ velocity fields for various stages of mergers
(UGC~993, SDSS~J1044+0353, SBS~1116+517, SBS~1159+545, HS~2236+1344).
In two cases we have good evidence for two independently rotating
discs (UGC~993 and HS~2236+1344).
\item
In the other galaxies, the rotation component of the overall velocity field is
important, but large disturbances appear. The residuals of
the `best-fitting' rotation model imply either a recent merger, or sufficiently
strong disturbance by nearby galaxies (HS~0822+3542 and UGC~772).
\item
Probable starburst-induced shells were identified in several
galaxies (UGC 772, UGC~993,  SBS~0335--052E, SBS~1116+517) through their
ionized gas velocity fields and velocity dispersion maps. We presented the
results of simple simulations of the expected velocity patterns which one
can observe in dwarf galaxies with expanding shells. To first order,
this allowed a feel for the effect of shells on the results of the
tilted-ring model.
\item
The interacting/merging nature of the binary system of the well separated XMD
galaxies SBS~0335--052E and W is best evident from their HI morphology
and kinematics data. Despite a relatively large mutual distance, the tidal
action of each component to the other clearly affects the gas dynamics of
these very gas-rich objects and triggers the current SF burst and very
likely the previous major SF episode  \citep[as emphasized by][]
{SBS0335_GMRT}.  To understand this unique interacting system
(a nice representative of high-redshift young galaxies) in more detail, one
needs a wide grid of models of interacting gas-rich galaxies, like, e.g.,
"Identikit"  \citep[][]{Barnes2009},   but including SF processes.
\item
The H$\alpha$ images  and velocity fields for two LSB dwarfs,
Anon~J012544+075957 and SAO~0822+3545, companions of two program XMD galaxies,
are obtained and analysed. They can be used in statistical studies of SF in
LSB dwarfs.  The star formation in SAO~0822+3545 is highly asymmetric and
takes place mainly in two knots at the northern edge, that is likely induced
by the recent interaction with the nearby XMD BCG HS~0822+3542.
\end{enumerate}

The statistics of XMD starbursting galaxies, for which kinematics of
gas were studied in detail, is still insufficient. Nevertheless, the results
of our FPI study of the ionized gas kinematics in the
subsample of nine XMD star-forming galaxies, along with the complementary
results of the GMRT HI study of a part of these and other XMD galaxies
\citep{Ekta08,SBS0335_GMRT,EC10},
indicate that strong interactions and mergers of very metal-poor
dwarf galaxies are one of the major or significant factors triggering
their current and recent starbursts. In particular, this is valid for
the six most metal-poor (12+$\log$(O/H)$<$7.30) dwarf starbursting
galaxies. Merger-induced starbursts are consistent with the idea that
the progenitors of
such rare objects either are old and have been evolving very slowly on the
cosmological  timescale before the current starburst have occurred, or they
are extremely  metal-poor because they are comparatively young, and thus
began their star formation and chemical enrichment with large delay.

\section*{Acknowledgements}

This research has made use of the NASA/IPAC Extragalactic Database (NED) which is operated by the Jet Propulsion Laboratory, California Institute of Technology, under contract with the National Aeronautics and Space Administration.
Funding for the Sloan Digital Sky Survey (SDSS) and SDSS-II has been provided by the Alfred P. Sloan Foundation, the Participating Institutions, the National Science Foundation, the US Department of Energy, the National Aeronautics and Space Administration, the Japanese Monbukagakusho, the Max Planck Society and the Higher Education Funding Council for England. The SDSS web site is http://www.sdss.org/. AVM and SAP acknowledge the support of this work through RFBR grant No. 06-02-16617.  AVM also acknowledges the support from RFBR grant No. 09-02-00870. AYK and SAP acknowledge support from the National Research Foundation of South Africa.
We thank B.~Ekta and J.~Chengalur for providing us with new information on HI morphology and velocity field of some of the studied XMD galaxies prior  publication, and A.~Burenkov - for the help with observations. We are grateful to S.~Crawford for the help in English improvement. The authors thank the anonymous referee for useful suggestions and questions, which helped to improve the paper.


\newpage

\newpage

\newpage

\appendix

\section[]{On the velocity field of disc with superimposed shell}

\label{App1}

The tilted-ring fit  is very popular for analysis of  gas velocity fields in
galactic discs (see \citet{Teuben2002} for detailed review). In
this approach, it is easy to connect the radial trends of kinematically
determined $PA$ and $i$ with real changes of disc geometrical parameters
(warp, polar rings, etc.) and/or with regular non-circular motions caused
by spiral arms or bar. However, in dwarf galaxies the SF-induced expanding
shells can significantly disturb the disc velocity field since the amplitude
of the rotation curve is rather small and the size of a shell can be
comparable
with the size of the disc. If we allow for variations of kinematical $PA$ with
radius, it can lead to a mistaken conclusion on disc orientation or about the
character of non-circular motions. In this Appendix we examine
the effect of
expanding shells on the tilted-ring fitting of artificial velocity fields.

The `observed' velocities in our two-component model are simply the sum of
line-of-sight velocities for the disc and the shell:
\begin{equation}
V_{obs}=V_{disc}+V_{shell} =V_{sys}+V_{rot}\cos\varphi\sin i+V_{shell}.
\label{eq1}
\end{equation}
This approximation  assumes the equal
contribution of both components in the
total velocity pattern in the overlapping regions, i.e., the same
brightness
of the components in each point.  Of course, this is not similar
to the real situation. However, we believe that eq.~(\ref{eq1}) can be used
as a first approximation. One of the constraints used in the models
below is that
$V_{shell}$ should not exceed the typical $FWHM$ of the observed emission
lines (about $30-50$~\kms).
Otherwise we would detect the two-component line profiles.

The azimuthal  angle in the plane of galaxy $\varphi$ is connected with the
position angle in the sky by the relation:
$\tan\varphi=\tan(PA-PA_0)/\cos~i$. The rising rotation curve is
constructed by the relation:
\begin{equation}
V_{rot}={{2}\over{\pi}}V_{max}\arctan{{R}\over{h}},
\end{equation}
with $h=1$ arcsec and the maximal velocity $V_{max}=30$~\kms\ which is
typical of dwarf galaxies in our  sample.

For the shell, we use a model of a thin expanding hemisphere. This
implies that we see only its approaching (blueshifted) side,
whereas the redshifted one is  obscured  by an interior extinction.
The model for the shell is:
\begin{equation}
V_{shell}=-V_{exp}\sqrt{1+{{(x-x_s)^2+(y-y_s)^2}\over{R_s^2}}}.
\end{equation}
Here $x_s$ and $y_s$ are sky-plane coordinates of the shell centre relative
to the disc centre. The shell has a radius  of $R_s=3.5$ arcsec and
the expansion velocity of $V_{exp}=$15~\kms.

Using eq.~(\ref{eq1}), we created models for moderately inclined  disc
($i=45^\circ$, $PA_0=45^\circ$) with the maximal radius
of 10 arcsec and with various values of $x_s$ and $y_s$.
A systemic velocity of $V_{sys}=1000$~\kms\ was adopted, that  is
representative of our sample. Its value has no effect on model isovelocity
patterns.

\begin{figure*}
\centerline{
\includegraphics[height=11 cm]{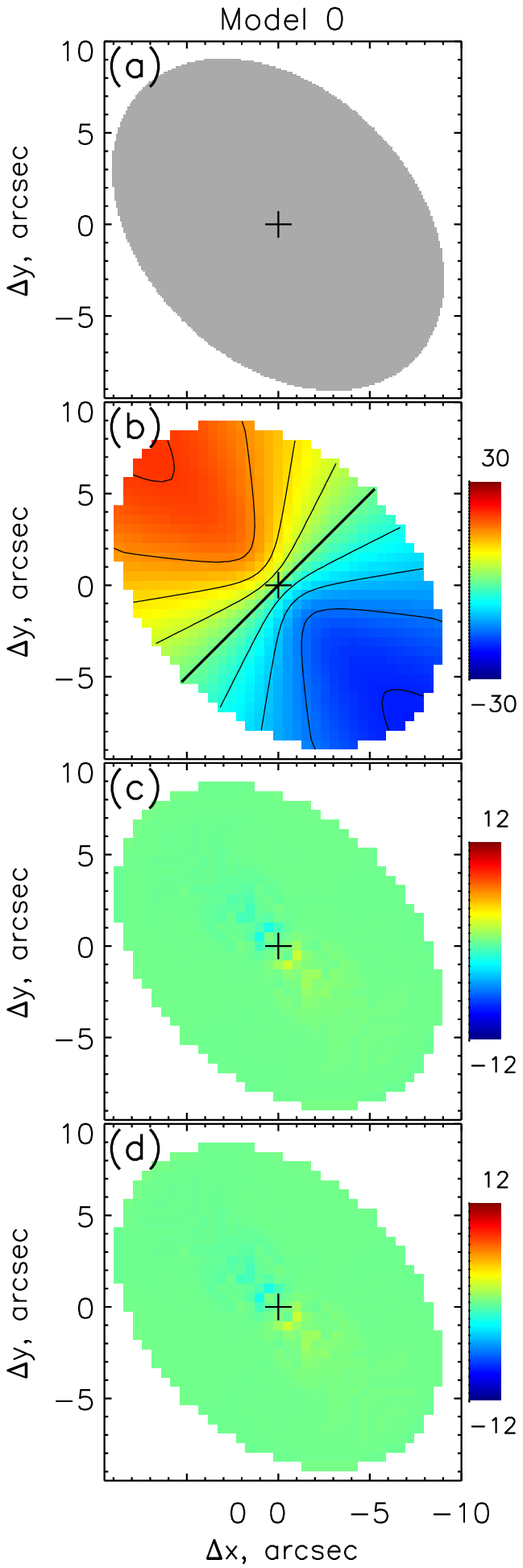}
\includegraphics[height=11 cm]{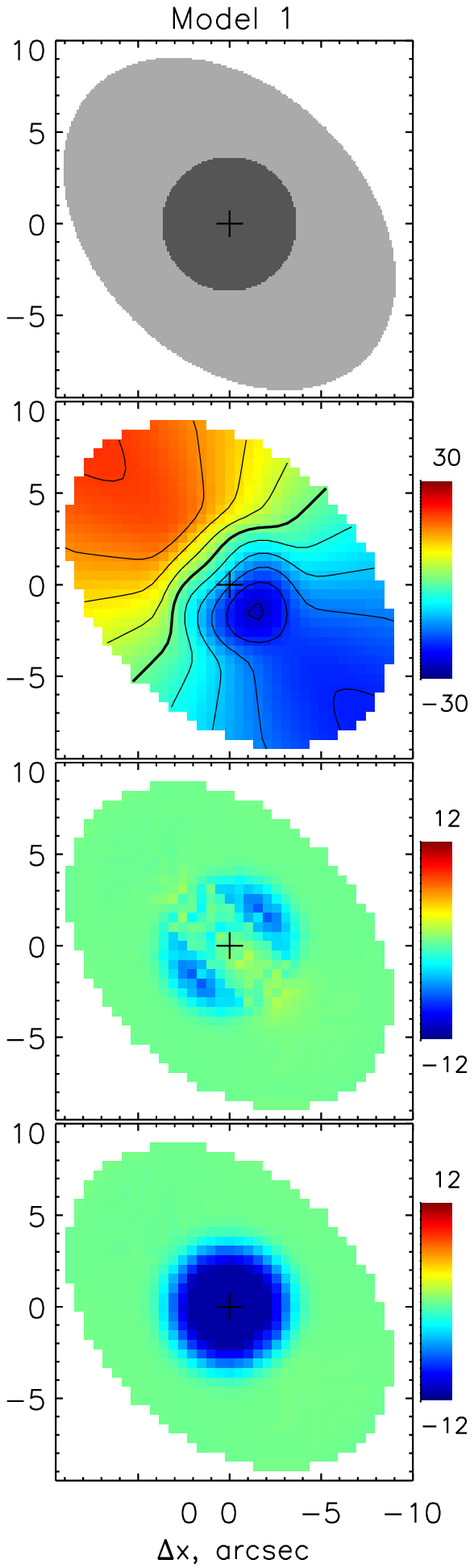}
\includegraphics[height=11 cm]{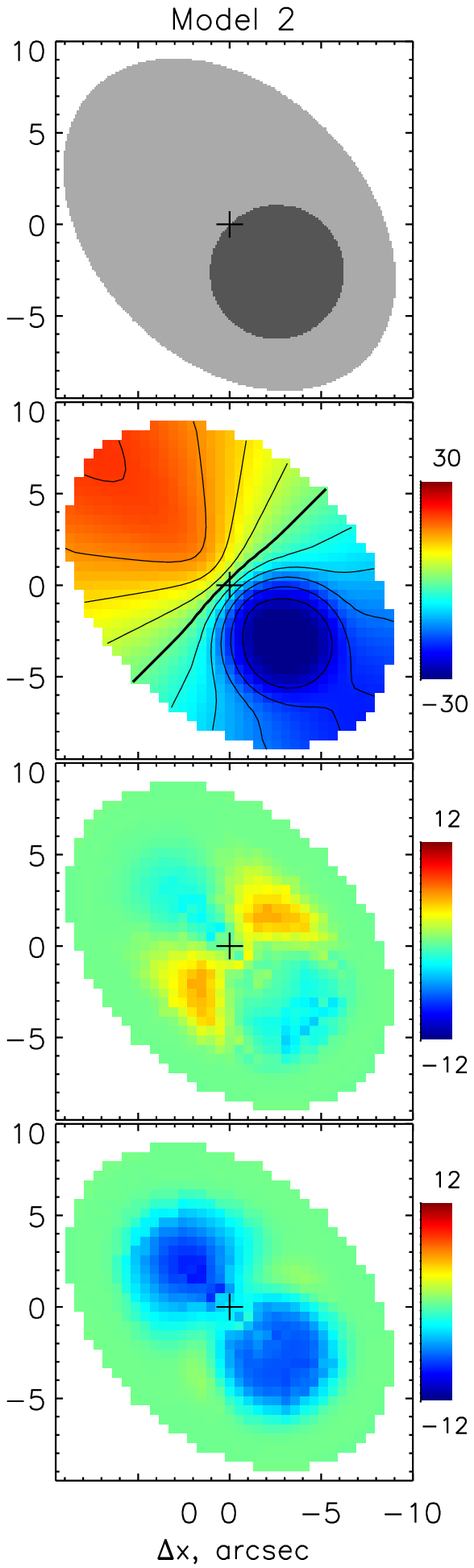}
\includegraphics[height=11 cm]{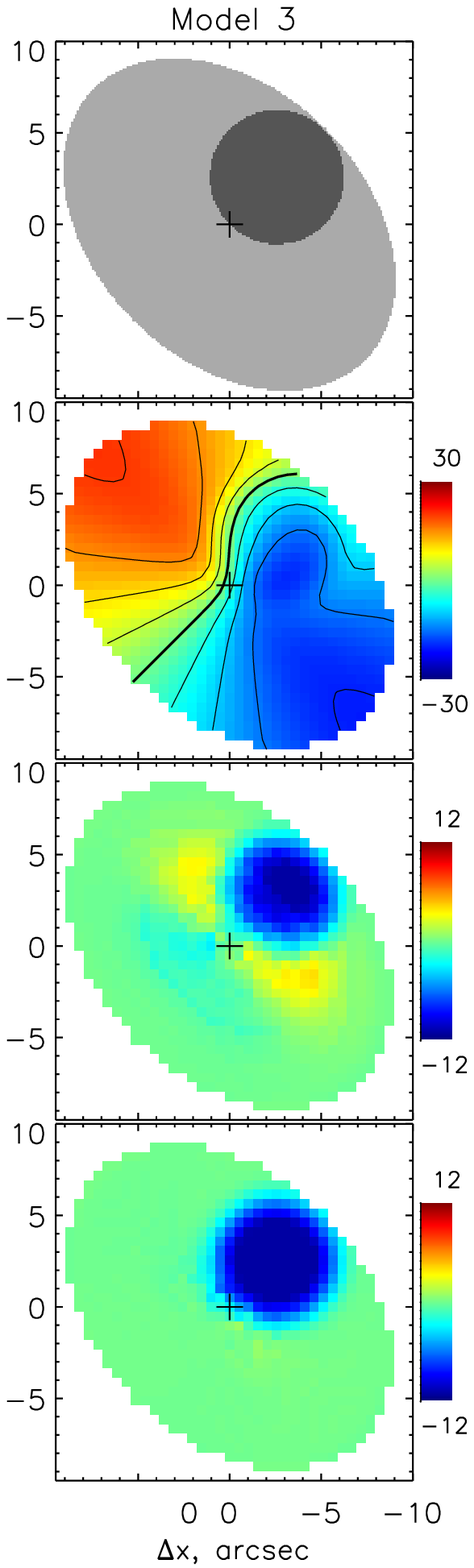}
\includegraphics[height=11 cm]{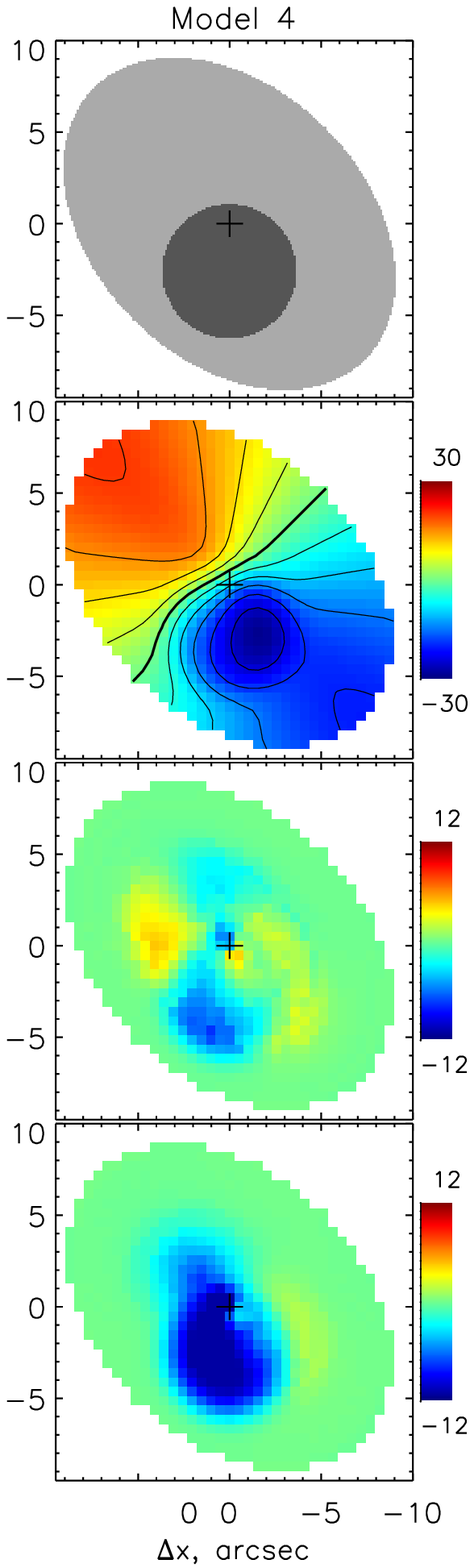}
}
\caption{The simulated maps. (a) -- the relative position of the disc (grey)
and shell (dark). (b) -- the simulated velocity field (after the subtraction
of $V_{sys}=1000$ \kms). (c) -- the residual velocity field after the
subtraction of the tilted-ring model (Method~I) which allows variations
of $PA$ and $V_{sys}$. (d) -- the same residual field, but for Method~II
($PA,V_{sys}=$const)}
\label{figA1}
\end{figure*}

 The models were
calculated on the grid with cell size 0.1 arcsec. Then the constructed
models were convolved with the 2D Gaussian with $FWHM=1.5$ arcsec that
simulated the typical seeing effect, and were rebinned with
the pixel size of 0.5 arcsec in accord with our observational conditions.
Fig.~\ref{figA1} shows examples of typical simulated velocity fields.

Model~0 represents  the disc without shell. In Model~1 the shell is
located in the disc centre. In Model~2 and Model~3, the centre of the shell
is shifted along the major and  the minor axes, respectively.
In Model~4 the shell centre is shifted in the direction of $PA=180$\degr.

\begin{figure*}
\centerline{
\includegraphics[height=8.5 cm]{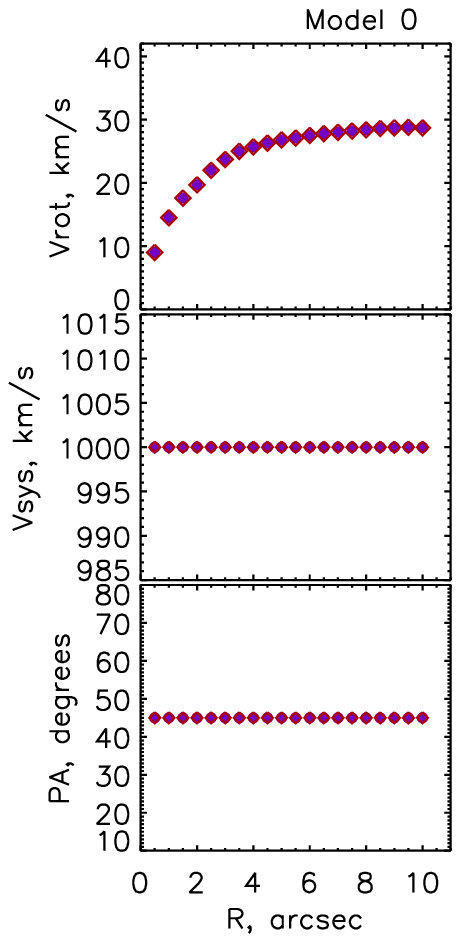}
\includegraphics[height=8.5 cm]{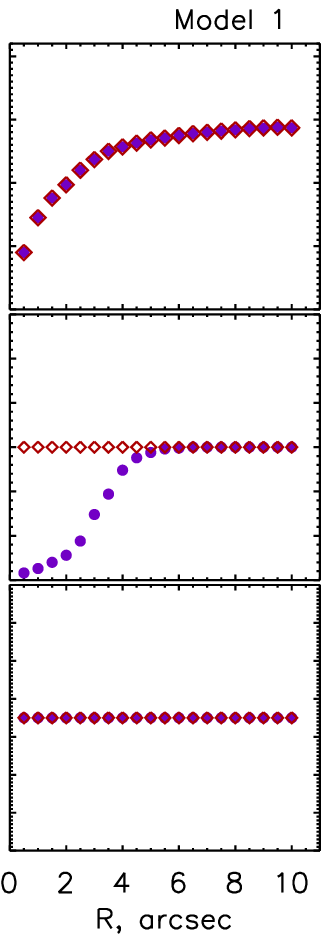}
\includegraphics[height=8.5 cm]{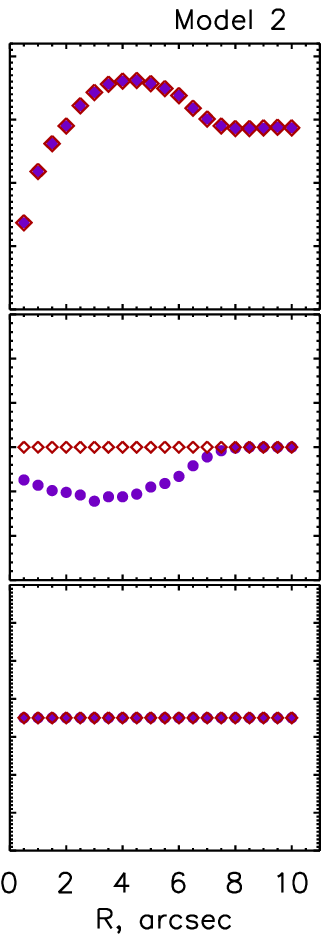}
\includegraphics[height=8.5 cm]{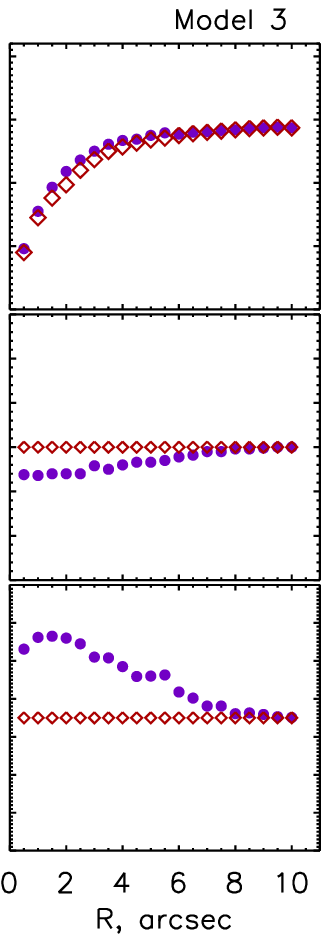}
\includegraphics[height=8.5 cm]{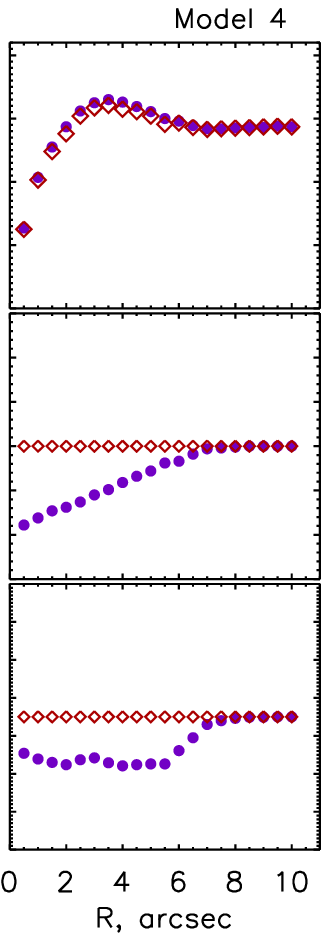}
}
\caption{The fitting parameters for the simulated velocity fields: rotation
velocity (top), position angle (middle) and systemic velocity (bottom).
Blue points and red diamonds show  results for Method~I and Method~II,
respectively.}
\label{figA2}
\end{figure*}

The simulated velocity fields were analysed with the same tilted-ring
model method under two approximations. Method~I allows for radial variations
of $PA$ and $V_{sys}$, while in Method~II these values were fixed and only
the rotation velocity was a free parameter. The inclination angle and the
rotation centre were fixed in both Methods. Fig.~\ref{figA1} shows the
residual velocities in the both approximations. The results of fitting are
shown in
Fig.~\ref{figA2}. These figures demonstrate how an expanding shell
affects  the observed velocity field and changes the estimates
of best-fitting  parameters. Namely, the central location of the shell, or its
offset along the disc major axis provoke the radial variations of $V_{sys}$
(see Models 1 and 2). The shell location beyond the disc major axis also
gives the systematic errors in the $PA$ estimations (Models 3 and 4). When
the shell centre lies on the minor axis, the radial trend of  $PA$ reach
the maximal amplitude  ($\sim$20\degr\ in our simulations). That large
variation of $PA$ can result in    erroneous conclusions concerning
the disc structure, like the presence of the inner warp or even of a nuclear
disc.

The residual velocities in the case of Method~I  seem spread over a larger
 portion
of the disc than the real area of the shell (Fig.~\ref{figA1}).
The residuals are relatively  small, usually about $\pm$5-7~\kms.
Moreover, the peak of negative (`blue') residuals usually
does not coincide with the centre of the shell.  In contrast, if $PA$
and $V_{sys}$ are constant
(i.e., for Method~II), the residual velocities pattern is in
a good agreement with the location of the shell, and the value of the
negative residuals  is consistent with  $V_{exp}$, with except of the
special case of Model~2.

It is worth noting that the rotation curve derived from the simulated
velocity field can be distorted independently of the analysis method,
as can be seen for
$V_{rot}$  plots for Models 2 and 4  in Fig.~\ref{figA2}. However, in
the Method~II approach,   we can unambiguously recognise from the
residual map
the perturbed portion of the disc and then  derive the real rotation
curve on the second iteration after masking the shell region.
The latter  is impossible if $PA$
and $V_{sys}$ are free parameters of the fitting, because in this case it
is difficult to recognise the shell location.

Based on the results presented above, we  have analysed the
velocity fields  of dwarf galaxies using the `tilted-ring' fitting
assuming that  $PA$ and $V_{sys}$ are constant with radius. When the first
approximation was done, we masked the regions with large residuals (usually,
larger than $5-10$~\kms) and repeated the fitting. The final model velocity
field  and the rotation curve, i.e., parameters of the disc orientation were
calculated after several such iterations.

\label{lastpage}

\end{document}